\documentclass[12pt]{article}
\usepackage{graphicx}
\usepackage{authblk}
\usepackage{orcidlink,amsmath,rotating,amssymb,amsthm,algorithm,epstopdf,multicol,multirow,gensymb,subcaption,fontenc,algorithmic,tablefootnote,longtable,array}
\usepackage[section]{placeins}
\usepackage{enumerate}
\usepackage{natbib}
\usepackage{url} 
\usepackage{hyperref}
\usepackage{xcolor,soul,chngcntr}
\hypersetup{colorlinks=true,linkcolor=,citecolor=blue}

\newcolumntype{C}[1]{>{\centering\arraybackslash}p{#1}}
\newcommand{\blind}{1}

\begin{document}

\def\spacingset#1{\renewcommand{\baselinestretch}%
{#1}\small\normalsize} \spacingset{1}


\if1\blind
{
  \title{\bf Robust semi-parametric mixtures of linear experts using the contaminated Gaussian distribution}
  \author{Peterson Mambondimumwe, Sphiwe B. Skhosana\textsuperscript{*}\orcidlink{0000-0002-4740-7540} and Najmeh Nakhaei Rad\orcidlink{0000-0002-7831-5614}\vspace{.5cm}\\
Department of Statistics, University of Pretoria, Pretoria, 0028, South Africa\vspace{.5cm}\\
    \textsuperscript{*}Corresponding author: \url{spiwe.skhosana@up.ac.za}}
  \maketitle
} \fi

\if0\blind
{
  \bigskip
  \bigskip
  \bigskip
  \begin{center}
    {\LARGE\bf Robust semi-parametric mixtures of linear experts using the contaminated Gaussian distribution}
\end{center}
  \medskip
} \fi

\bigskip
\begin{abstract}
Semi- and non-parametric mixture of regressions are a very useful flexible class of mixture of regressions in which some or all of the parameters are non-parametric functions of the covariates. These models are, however, based on the Gaussian assumption of the component error distributions. Thus, their estimation is sensitive to outliers and heavy-tailed error distributions. In this paper, we propose semi- and non-parametric contaminated Gaussian mixture of regressions to robustly estimate the parametric and/or non-parametric terms of the models in the presence of mild outliers. The virtue of using a contaminated Gaussian error distribution is that we can simultaneously perform model-based clustering of observations and model-based outlier detection. We propose two algorithms, an expectation-maximization (EM)-type algorithm and an expectation-conditional-maximization (ECM)-type algorithm, to perform maximum likelihood and local-likelihood kernel estimation of the parametric and non-parametric of the proposed models, respectively. The robustness of the proposed models is examined using an extensive simulation study. The practical utility of the proposed models is demonstrated using real data.
\end{abstract}

\noindent%
\vfill

\newpage
\spacingset{1.4} 

\section{Introduction}\label{sec1} 
Finite mixtures of linear regression models are very useful for studying the relationship between a response variable $y$ and a set of covariates $\mathbf{x}=(x_1,x_2,\dots,x_p)^\top$ when the underlying population is made up of a, typically unknown, number of, say $K$, unobserved sub-populations also known as components.\\
Let $Z$ be a component indicator variable with a discrete distribution $P(Z=k)=\pi_k$, for $k=1,2,\dots,K$, where each $\pi_k>0$ and $\sum^{K}_{k=1}\pi_k=1$. Given that $Z=k$ and $\mathbf{X}=\mathbf{x}$, the relationship between $y$ and $\mathbf{x}$ is given by
\begin{equation}
	y=\mathbf{x}^\top\boldsymbol{\beta}_k+\epsilon_k,\text{ for } k=1,2,\dots,K,
\end{equation}
where $\mathbf{x}=(1,x_1,x_2,\dots,x_p)^\top$, $\boldsymbol{\beta}_k=(\beta_{k0},\beta_{k1},\dots,\beta_{kp})$ is the regression parameter vector, $\epsilon_k$ is the error term, typically assumed to: (i) be independent of $\mathbf{x}$ and (ii) follow a Gaussian distribution with zero mean and variance $\sigma^2_k$. That is $\epsilon_k\sim \mathcal{N}(0,\sigma^2_k)$, where $\mathcal{N}(\mu,\sigma^2)$ denotes the Gaussian distribution with mean $\mu$ and variance $\sigma^2$. Since $Z$ is typically unobserved, given only $\mathbf{X}=\mathbf{x}$, $y$ is said to follow a mixture of Gaussian distributions with a conditional density function
\begin{eqnarray}
	f(y|\mathbf{X}=\mathbf{x})&=&\sum_{k}^K\pi_k\mathcal{N}\{y|\mathbf{x}^\top\boldsymbol{\beta}_k,\sigma^2_k\},\label{model1}
\end{eqnarray}
where the weights $\pi_k$'s are the mixing probabilities specifying the relative size of each component. Model \eqref{model1} is the Gaussian mixture of linear regressions (GMLRs).\\
Although flexible, the applicability of the mixtures of linear regressions \eqref{model1} is limited by the assumption that the mixing proportions are constants. To increase the capability of the model, \cite{jacobs1991} proposed the mixture of experts (MoE) by allowing the mixing proportions to be functions of a set of covariates.\\
Let $\mathbf{t}=(1,t_1,t_2,\dots,t_q)^\top$ be a set of covariates, the MoE uses the multinomial logistic function to model the mixing proportion as a function of $\mathbf{t}$
\begin{eqnarray}
	\pi_k(\mathbf{t}|\boldsymbol{\gamma})=P(Z=k|\mathbf{t})=\frac{\text{exp}(\mathbf{t}^\top\boldsymbol{\gamma}_k)}{\sum^K_{j=1}\text{exp}(\mathbf{t}^\top\boldsymbol{\gamma}_j)},
\end{eqnarray}
where $\boldsymbol{\gamma}=(\boldsymbol{\gamma}_1,\boldsymbol{\gamma}_2,\dots,\boldsymbol{\gamma}_{K-1})$, with $\boldsymbol{\gamma}_k=(\gamma_{k0},\gamma_{k1},\dots,\gamma_{kq})$ as the coefficient vector for the $k^{th}$ component and $\boldsymbol{\gamma}_{K}$ is the null vector for identifiability.\\
Given $\mathbf{x}$ and $\mathbf{t}$, the conditional mixture density of $y$ is 
\begin{eqnarray}
	f(y|\mathbf{X}=\mathbf{x},\mathbf{T}=\mathbf{t})&=&\sum_{k}^K\pi_k(\mathbf{t}|\boldsymbol{\gamma})\mathcal{N}\{y|\mathbf{x}^\top\boldsymbol{\beta}_k,\sigma^2_k\}\label{model2},
\end{eqnarray}
where $\pi_k(\mathbf{t}|\boldsymbol{\gamma})>0$ and $\sum^K_{k=1}\pi_k(\mathbf{t}|\boldsymbol{\gamma})=1$ for all $\mathbf{t}$.\\
Model \eqref{model2} is the Gaussian MoE (GMoE) model in which the mixing proportions $\pi_k(\mathbf{t}|\boldsymbol{\gamma})$ are known as the gating functions and the component densities $\mathcal{N}\{y|\mathbf{x}^\top\boldsymbol{\beta}_k,\sigma^2_k\}$ are known as experts. Note that the covariates $\mathbf{x}$ and $\mathbf{t}$ may be identical.\\
The MoE \eqref{model2} is still limited by the assumption that the gating functions $\pi_k(\mathbf{t}|\boldsymbol{\gamma})$ are multinomial logistic functions. The monotonic shape of the logistic function may not be appropriate (\cite{young2010}). To further increase the flexibility of the MoE model, \cite{young2010} and \cite{huang2012} proposed to model the gating functions non-parametrically
\begin{eqnarray}
	f(y|\mathbf{X}=\mathbf{x},\mathbf{T}=\mathbf{t})&=&\sum_{k}^K\pi_k(\mathbf{t})\mathcal{N}\{y|\mathbf{x}^\top\boldsymbol{\beta}_k,\sigma^2_k\}\label{model3},
\end{eqnarray}
where $\pi_k(\mathbf{t})$ is a smooth unknown function of the covariates $\mathbf{t}$, hence non-parametric.\\ 
Consider a random sample $\{(\mathbf{x}_i,\mathbf{t}_i,y_i):i=1,2,\dots,n\}$,
\cite{young2010} proposed to model the non-parametric gating function as $\pi_k(\mathbf{t}_i)=\mathbb{E}\{Z_{ik}|\mathbf{t}_i\}$, where $Z_{ik}=1$ if the $i^{th}$ data point was generated by the $k^{th}$ expert and $0$ otherwise. \cite{young2010} used the local-polynomial regression approach to obtain a non-parametric estimate of the gating function by taking $Z_{ik}$ as the output of the fitted GMLRs \eqref{model1}. \cite{huang2012} suggested to model the non-parametric gating function fully non-parametrically using the local-likelihood approach (\cite{tibshirani1987}). The above non-parametric estimation procedures cannot handle high-dimensional covariates. To address this limitation, \cite{xiang2020} proposed to model the non-parametric gating function using a single-index model $\pi_k(\mathbf{t}^\top_i\boldsymbol{\gamma}_k)$, where $\boldsymbol{\gamma}_k$ is a single-index parameter vector. As an alternative, \cite{xue2022} proposed to model the non-parametric gating function using a neural network.\\
Note that the models discussed so far assume that each component follows a Gaussian distribution. Unfortunately, in practice, data may be contaminated by outliers, which may have a significant negative effect on the model estimation \cite{mazza2020}. \citep{mazza2020} distinguished between mild and gross outliers. The authors define mild outliers as points that do not significantly deviate from the regression relationship within a given component; however, these points produce component distributions that are too heavy-tailed to be adequately modelled by a Gaussian distribution. On the other hand, gross outliers are points that are significantly away from any of the regression components. Our focus in this paper is on mild outliers. For more details about the distinction between these two classes of outliers, see \cite{mazza2020} and the references therein.
For Gaussian mixtures of regressions, the most popular approach used to protect against outliers is to replace each Gaussian component with a heavy-tailed component distribution \cite{mazza2020}. \cite{yao2014} proposed robust mixtures of regressions by assuming that each component follows a t-distribution (hereafter, t-MR). \cite{song2014} made a similar robust proposal; however, they assumed that each component follows a Laplace distribution (hereafter L-MR). More recently, \cite{mazza2020} used the contaminated Gaussian distribution to model the distribution of each component of a mixture of regression models (hereafter CG-MLR). A more general approach, which also accounts for asymmetry, was proposed by \cite{zeller2016} by modelling each component using the scale-mixture of skew Gaussian distribution (hereafter SMSG-MR). However, the robust mixtures of regression models of \cite{yao2014}, \cite{song2014}, and \cite{mazza2020} assume that the mixing proportions are constants. \cite{chamroukhi2016}, \cite{nguyen2016}, and \cite{mirfarah2021} proposed robust mixtures of experts (MoE) models by extending the models of \cite{yao2014}, \cite{song2014}, and \cite{mazza2020}, respectively, to model the mixing proportions as functions of a set of covariates. The former three models are hereafter referred to as t-MoE, L-MoE, and CG-MoE, respectively. Note that the CG-MoE is a special case of the model proposed by \cite{mirfarah2021}. However, the model is too general for our objectives in this paper.\\
Note that the t-MoE, L-MoE, and the CG-MoE use the multinomial logistic function to model the gating functions or mixing proportions. Thus, their practical applicability is limited by this parametric assumption. An obvious and simple approach to enhance the flexibility of the robust MoLE models is to model the gating functions non-parametrically. However, currently, there is little attention given to this line of research. The only exception is \cite{ge2024}, who proposed to extend the t-MoLE model by assuming that the gating functions are non-parametric functions of the covariates. The model is hereafter referred to as the S-t-MoLE. The authors demonstrated the robustness and practical utility of the S-t-MoLE model when the component distribution has heavier tails than the Gaussian distribution.\\ 
In this paper, we propose a robust semi-parametric mixture of linear experts by assuming that each component distribution follows a contaminated Gaussian distribution. The resulting model is hereafter referred to as the S-CG-MoLE. In contrast to the S-t-MoLE, the proposed S-CG-MoLE has a principled and probabilistic approach for simultaneous clustering of observations and detection of mild outliers. Once the S-CG-MoLE model is fitted to the data, each observation can be first assigned to one component or cluster and thereafter classified as being a typical or atypical (outlier) observation (see \cite{mazza2020}). This paper makes the following contributions:
\begin{enumerate}
	\item We propose a robust and flexible mixture of regression models by assuming that each component follows a contaminated Gaussian distribution and the mixing proportion functions are non-parametric functions of the covariates.
	\item We introduce a modified Expectation-Conditional-Maximization (ECM) algorithm for simultaneous parametric and non-parametric maximum likelihood estimation.
\end{enumerate}
The remainder of the paper is structured as follows. In Section \ref{sec3}, we review the contaminated Gaussian mixture of linear experts (CG-MoLE) and its estimation procedure (summarized in Algorithm \ref{Alg1}) based on the expectation-conditional maximization (ECM) algorithm. This is followed by an extension of the CG-MoLE to the proposed semi-parametric CG-MoLE model with varying non-parametric mixing proportions, and thereafter we introduce its proposed estimation procedure (summarized in Algorithm \ref{Alg2}). We end the section by discussing model selection for and model-based clustering using the proposed model, respectively. In Section \ref{sec4}, conduct a simulation study to evaluate the robustness of the proposed model under various scenarios. In Section \ref{sec5}, we apply the proposed model to a real dataset. Finally, in Section \ref{sec6}, we conclude the paper and give directions for future studies.
\section{Methodology}\label{sec3}
\subsection{Contaminated Gaussian mixture of linear experts }
In this section, we define the CG-MoLE model and its estimation using the maximum likelihood method.
\subsubsection*{Model definition}
The CG-MoLE model, obtained as a special case of the model in \cite{mirfarah2021}, is an extension of the GMoLE model. The form of the $K-$component CG-MoLE model is
\begin{eqnarray}\label{CG-MoLE}
	f(y|\mathbf{X}=\mathbf{x},\mathbf{T}=\mathbf{t})=\sum^{K}_{k=1}\pi_k(\mathbf{t}|\boldsymbol{\gamma})\mathcal{CN}\{y|\mathbf{x};\boldsymbol{\theta}_k\},
\end{eqnarray}
where 
\begin{eqnarray}\label{CG_pdf}
	\mathcal{CN}\{y|\mathbf{x};\boldsymbol{\theta}_k\}=\alpha_k\mathcal{N}\{y|\mathbf{x}^\top\boldsymbol{\beta}_k,\sigma^2_k\}+(1-\alpha_k)\mathcal{N}\{y|\mathbf{x}^\top\boldsymbol{\beta}_k,\eta_k\sigma^2_k\}
\end{eqnarray}
is the density function of the contaminated Gaussian distribution for the $k^{th}$ component, and $\boldsymbol{\theta}_k=(\alpha_k,\boldsymbol{\beta}_k,\eta_k,\sigma^2_k)$ is the parameter vector of the $k^{th}$ contaminated Gaussian expert. The parameter $\alpha_k\in (0,1)$ represents the proportion of typical (uncontaminated) data points and $\eta_k>1$ is the degree of contamination. The latter parameter measures the extent to which the variance has been inflated by the presence of outliers. \\
Let $\boldsymbol{\Theta}=(\boldsymbol{\gamma}_1,\boldsymbol{\gamma}_2,\dots,\boldsymbol{\gamma}_K;\boldsymbol{\theta}_1,\boldsymbol{\theta}_2,\dots,\boldsymbol{\theta}_K)$ be the parameter vector of the CG-MoLE model. The CG-MoLE model parameters can be estimated using maximum likelihood estimation. We turn to this next.
\subsubsection*{Model estimation}
Consider a random sample $\{(\mathbf{x}_i,\mathbf{t}_i,y_i):i=1,2,\dots,n\}$ from the CG-MoLE model \eqref{CG-MoLE}. The corresponding observed log-likelihood function is 
\begin{eqnarray}
	\ell\{\boldsymbol{\Theta}\}=\sum_{i=1}^n\text{log}\sum_{k=1}^K\pi_k(\mathbf{t}_i|\boldsymbol{\gamma})[\alpha_k\mathcal{N}\{y|\mathbf{x}^\top\boldsymbol{\beta}_k,\sigma^2_k\}+(1-\alpha_k)\mathcal{N}\{y|\mathbf{x}^\top\boldsymbol{\beta}_k,\eta_k\sigma^2_k\}].
\end{eqnarray}
The MLE of $\boldsymbol{\Theta}$ is
\begin{eqnarray}\label{MLE1}
	\hat{\boldsymbol{\Theta}}_{\text{MLE}}=\max_{\boldsymbol{\Theta}}\ell\{\boldsymbol{\Theta}\}.
\end{eqnarray}
The expression \eqref{MLE1} has no closed-form solution due to the presence of missing data. More specifically, there are two sources of missing data. The first source of missing data arises from the fact that we don't know which component generated which data point. The second source of missing data arises from the fact that we don't know whether a data point in the $k^{th}$ component is an outlier or not an outlier. Therefore, the MLE of $\boldsymbol{\Theta}$ is obtained by maximizing the observed log-likelihood function $\ell\{\boldsymbol{\Theta}\}$ using the Expectation-Conditional-Maximization (ECM) algorithm \cite{meng1993}. The ECM is an extension of the traditional Expectation-Maximization (EM) algorithm \cite{DLR1977} for maximizing the observed log-likelihood in the presence of missing data. The ECM iterates between three steps: an Expectation (E-) step and two Conditional Maximization (CM-) steps until convergence. The two CM-steps arise as a result of the partition $\boldsymbol{\Theta}=\{\boldsymbol{\Theta}_1,\boldsymbol{\Theta}_2\}$, where $\boldsymbol{\Theta}_1=\{\alpha_k,\boldsymbol{\beta}_k,\sigma^2_k,\boldsymbol{\gamma}_k\}_{k=1}^K$ and $\boldsymbol{\Theta}_2=\{\eta_k\}_{k=1}^K$. The E-step computes the expected complete-data log-likelihood, and the CM-steps maximize it. The complete-data is $\{(\mathbf{x}_i,\mathbf{r}_i,y_i,\mathbf{z}_i,\mathbf{v}_i):i=1,2,\dots,n\}$, where $\mathbf{z}_i=(z_{i1},\dots,z_{iK})$, with $z_{ik}=1$ if the $i^{th}$ data point comes from the $k^{th}$ component and $0$ otherwise, and $\mathbf{v}_i=(v_{i1},\dots,\mathbf{v}_{iK})$, with $v_{ik}=1$ if the $i^{th}$ data point in the $k^{th}$ component is good and $v_{ik}=0$ otherwise. The log-likelihood based on the complete-data is
\begin{eqnarray}\label{llc1}
	\ell_c\{\boldsymbol{\Theta}\}=\ell_{1c}\{\boldsymbol{\gamma}\}+\ell_{2c}\{\boldsymbol{\alpha}\}+\ell_{3c}\{\boldsymbol{\vartheta}\},
\end{eqnarray}
where
\begin{eqnarray}
	\ell_{1c}\{\boldsymbol{\gamma}\}&=&\sum_{i=1}^n\sum_{k=1}^Kz_{ik}\text{ln}\pi_k(\mathbf{t}_i|\boldsymbol{\gamma}),\nonumber\\
	\ell_{2c}\{\boldsymbol{\alpha}\}&=&\sum_{i=1}^n\sum_{k=1}^Kz_{ik}[v_{ik}\text{ln}\alpha_k+(1-v_{ik})\text{ln}(1-\alpha_k)],\nonumber\\
	\ell_{3c}\{\boldsymbol{\vartheta}\}&=&-\frac{1}{2}\sum_{i=1}^n\sum_{k=1}^K\bigg[z_{ik}\text{ln}\sigma^2_k+z_{ik}(1-v_{ik})\text{ln}\eta_k+z_{ik}\bigg(1+\frac{1-v_{ik}}{\eta_k}\bigg)\frac{(y_i-\mathbf{x}^\top_i\boldsymbol{\beta}_k)^2}{\sigma^2_k}\bigg],\nonumber\\
\end{eqnarray}
with $\boldsymbol{\alpha}=(\alpha_1,\alpha_2,\dots,\alpha_K)$ and $\boldsymbol{\vartheta}=(\boldsymbol{\beta}_1,\boldsymbol{\beta}_2,\dots,\boldsymbol{\beta}_K;\sigma^2_1,\sigma^2_2,\dots,\sigma_K^2;\eta_1,\eta_2,\dots,\eta_K)$.
\paragraph{E-step}
In the E-step, at the $r^{th}$ iteration of the ECM algorithm, we calculate the conditional expectation of $\ell_c\{\boldsymbol{\Theta}\}$, denoted $Q\{\boldsymbol{\Theta|\boldsymbol{\Theta}^{(r-1)}}\}$. This reduces to calculating $\mathbb{E}\{z_{ik}|y_i,\mathbf{x}_i,\mathbf{t}_i\}$ and $\mathbb{E}\{v_{ik}|y_i,\mathbf{x}_i,\mathbf{t}_i,\mathbf{z}_i\}$, for $i=1,2,\dots,n$ and $k=1,2,\dots,K$, respectively, as

\begin{eqnarray}
	\mathbb{E}\{z_{ik}|y_i,\mathbf{x}_i,\mathbf{t}_i\}\equiv z_{ik}^{(r)}&=&\frac{\pi_{k}(\mathbf{t}_i|\boldsymbol{\gamma}^{(r-1)})\mathcal{CN}\{y_i|\mathbf{x}_i;\boldsymbol{\theta}^{(r-1)}_k\}}{\sum_{\ell=1}^K\pi_{\ell}(\mathbf{t}_i|\boldsymbol{\gamma}^{(r-1)})\mathcal{CN}\{y_i|\mathbf{x}_i;\boldsymbol{\theta}^{(r-1)}_\ell\}}\label{resp1},\\
	\mathbb{E}\{v_{ik}|y_i,\mathbf{x}_i,\mathbf{t}_i,\mathbf{z}_i\}\equiv v^{(r)}_{ik}&=&\frac{\alpha^{(r-1)}_k\mathcal{N}\{y_i|\mathbf{x}^\top_i\boldsymbol{\beta}^{(r-1)}_k,\sigma^{2(r-1)}_k\}}{\mathcal{CN}\{y_i|\mathbf{x}_i;\boldsymbol{\theta}^{(r-1)}_k\}}.\label{resp2}
\end{eqnarray}
Equation \eqref{resp1} can be interpreted as the posterior probability that the $i^{th}$ data point belongs to the $k^{th}$ component, whereas \eqref{resp2} can be interpreted as the posterior probability that the $i^{th}$ data point is not an outlier in component $k$. Substituting $z_{ik}$ with $z_{ik}^{(r)}$ and $v_{ik}$ with $v^{(r)}_{ik}$ in \eqref{llc1}, we obtain
\begin{eqnarray}\label{expllc1}
	Q\{\boldsymbol{\Theta}|\boldsymbol{\Theta}^{(r-1)}\}=Q\{\boldsymbol{\gamma}|\boldsymbol{\gamma}^{(r-1)}\}+Q\{\boldsymbol{\alpha}|\boldsymbol{\alpha}^{(r-1)}\}+Q\{\boldsymbol{\vartheta}|\boldsymbol{\vartheta}^{(r-1)}\},
\end{eqnarray}
where 
\begin{eqnarray}
	Q\{\boldsymbol{\gamma}|\boldsymbol{\gamma}^{(r-1)}\}&=&\sum_{i=1}^n\sum_{k=1}^Kz^{(r)}_{ik}\text{ln}\pi_k(\mathbf{t}_i|\boldsymbol{\gamma}),\nonumber\\
	Q\{\boldsymbol{\alpha}|\boldsymbol{\alpha}^{(r-1)}\}&=&\sum_{i=1}^n\sum_{k=1}^Kz^{(r)}_{ik}[v^{(r)}_{ik}\text{ln}\alpha_k+(1-v^{(r)}_{ik})\text{ln}(1-\alpha_k)],\nonumber\\
	Q\{\boldsymbol{\vartheta}|\boldsymbol{\vartheta}^{(r-1)}\}&=&-\frac{1}{2}\sum_{i=1}^n\sum_{k=1}^K\bigg[z^{(r)}_{ik}\text{ln}\sigma^2_k+z^{(r)}_{ik}(1-v^{(r)}_{ik})\text{ln}\eta_k+z^{(r)}_{ik}\bigg(1+\frac{1-v^{(r)}_{ik}}{\eta_k}\bigg)\frac{(y_i-\mathbf{x}^\top_i\boldsymbol{\beta}_k)^2}{\sigma^2_k}\bigg].\nonumber\\
\end{eqnarray}

\paragraph{CM-step 1}
In the first CM-step, at the $r^{th}$ iteration, of the ECM algorithm, we calculate $\boldsymbol{\Theta}^{(r)}_1$, by maximizing $Q\{\boldsymbol{\Theta}|\boldsymbol{\Theta}^{(r-1)}\}$ with $\boldsymbol{\Theta}_2$ fixed at $\boldsymbol{\Theta}^{(r-1)}_2$. In particular, by maximizing $Q\{\boldsymbol{\alpha}|\boldsymbol{\alpha}^{(r-1)}\}$ with respect to $\alpha_k$, for $k=1,2,\dots,K$, and $Q\{\boldsymbol{\vartheta}|\boldsymbol{\vartheta}^{(r-1)}\}$ with respect to $\boldsymbol{\beta}_k$ and $\sigma^2_k$, for $k=1,2,\dots,K$, respectively, we obtain
\begin{eqnarray}
	\alpha^{(r)}_k&=&\frac{1}{n^{(r)}_k}\sum_{i=1}^nz^{(r)}_{ik}v^{(r)}_{ik}\label{alpha_est},\\
	\boldsymbol{\beta}^{(r)}_k&=&(\mathbf{X}^\top \mathbf{W}^{(r)}_k\mathbf{X})^{-1}\mathbf{X}^\top \mathbf{W}^{(r)}_k\mathbf{y}\label{beta_est},\\
	\sigma^{2(r)}_k&=&\frac{1}{n^{(r)}_k}\sum_{i=1}^n w^{(r)}_{ik}(y_i-\mathbf{x}^\top_i\boldsymbol{\beta}^{(r)}_k)^2,\label{sigma_est}
\end{eqnarray}
where $n^{(r)}_k=\sum_{i=1}^nz^{(r)}_{ik}$, $\mathbf{X}=(\mathbf{x}^\top_1,\mathbf{x}^\top_2,\dots,\mathbf{x}^\top_n)^\top$ is the $n\times (p+1)$ design matrix, $\mathbf{y}=(y_1,y_2,\dots,y_n)$ and $\mathbf{W}^{(r)}_k=\text{diag}\{w^{(r)}_{1k},w^{(r)}_{2k},\dots,w^{(r)}_{nk}\}$ is an $n\times n$ diagonal matrix, with $w^{(r)}_{ik}=z^{(r)}_{ik}\bigg(v^{(r)}_{ik}+\frac{1-v^{(r)}_{ik}}{\eta^{(r-1)}_k}\bigg)$, for $i=1,2,\dots,n$ and $k=1,2,\dots,K$.\\
The maximization of $Q\{\boldsymbol{\gamma}|\boldsymbol{\gamma}^{(r-1)}\}$ with respect to $\boldsymbol{\gamma}_k$, for $k=1,2,\dots,K$, does not have a closed-form expression. This is typically performed numerically using the Newton-Raphson algorithm \cite{bishop2006}.\\
Given the initial value $\boldsymbol{\gamma}^{(0)}_k$, at the $s^{th}$ iteration of the Newton-Raphson algorithm, we update $\boldsymbol{\gamma}^{(s)}_k$ as follows
\begin{eqnarray}\label{NR}
	\boldsymbol{\gamma}^{(s)}_k=\boldsymbol{\gamma}^{(s-1)}_k-\mathbf{H}^{-1}[\boldsymbol{\gamma}^{(s-1)}_k]\mathbf{g}[\boldsymbol{\gamma}^{(s-1)}_k],
\end{eqnarray}
where $\mathbf{g}[\boldsymbol{\gamma}^{(s-1)}_k]=\frac{\partial Q\{\boldsymbol{\gamma}|\boldsymbol{\gamma}^{(r-1)}\}}{\partial \boldsymbol{\gamma}_k}$ and $\mathbf{H}[\boldsymbol{\gamma}^{(s-1)}_k]=\frac{\partial^2 Q\{\boldsymbol{\gamma}|\boldsymbol{\gamma}^{(r-1)}\}}{\partial^2\mathbf{\gamma}_k}$ is the gradient vector and Hessian matrix of $Q\{\boldsymbol{\gamma}|\boldsymbol{\gamma}^{(r-1)}\}$, respectively. The value of $\boldsymbol{\gamma}^{(r)}_k$ is taken as the value of $\boldsymbol{\gamma}_k$, for $k=1,2,\dots,K$, at convergence of the Newton-Raphson algorithm in \eqref{NR}. 

\paragraph{CM-step 2}
In the second CM-step, at the $r^{th}$ iteration of the ECM algorithm, we calculate $\boldsymbol{\Theta}^{(r)}_2$ by maximizing $Q\{\boldsymbol{\Theta}|\boldsymbol{\Theta}^{(r-1)}\}$ with $\boldsymbol{\Theta}_1$ fixed at $\boldsymbol{\Theta}^{(r)}_1$ obtained at CM-step 1. In particular, we maximize 
\begin{eqnarray}\label{eta_obj1}
	-\frac{1}{2}\sum_{i=1}^n z_{ik}^{(r)}(1-v^{(r)}_{ik})\text{ln}\eta_k-\frac{1}{2}\sum_{i=1}^nz^{(r)}_{ik}\frac{(1-v^{(r)}_{ik})}{\eta_k}\frac{(y_i-\mathbf{x}^\top_i\boldsymbol{\beta}^{(r)}_k)^2}{\sigma^{2(r)}_k}
\end{eqnarray}
with respect to $\eta_k$, given the constraint $\eta_k>1$, for $k=1,2,\dots,K$.\\
The above ECM algorithm is summarized in Algorithm \ref{Alg1}. 
\begin{algorithm}
	\caption{ECM Algorithm for fitting CG-MoLE}
	\label{Alg1}
	Let $r = 0$ represent the current iteration number, $\boldsymbol{\theta}_k=(\boldsymbol{\beta}_{k},\sigma^{2}_{k},\alpha_{k},\eta_{k})$ and $\boldsymbol{\psi}_k=( \boldsymbol{\gamma}_k,\boldsymbol{\beta}_{k},\sigma^2_{k},\alpha_{k},\eta_{k}\ )$. The algorithm alternates between the E-step and two CM-steps until convergence.
	\begin{enumerate}
		\item[] \textbf{E-step}: Computing the posterior probabilities $z^{(r)}_{ik}$ and $v^{(r)}_{ik}$, for $i=1,2,\dots,n$ and $k=1,2,\dots,K$, using \eqref{resp1} and \eqref{resp2}, respectively.\\
		
		\item[] \textbf{CM-step 1}: Updating the parameters $\alpha_k$, $\boldsymbol{\beta}_k$, $\sigma^2_k$, for $k=1,2,\dots,K$, and $\boldsymbol{\gamma}_k$, for $k=1,2,\dots,K-1$. 
		\begin{itemize}
			\item Given $\eta^{(r-1)}_k$, calculating $\alpha^{(r)}_k$, $\boldsymbol{\beta}^{(r)}_k$ and $\sigma^{2(r)}_k$ using \eqref{alpha_est}, \eqref{beta_est} and \eqref{sigma_est}.
			\item Calculating $\boldsymbol{\gamma}^{(r)}_k$, by iterating the Newton-Raphson update \eqref{NR} for $\boldsymbol{\gamma}^{(s)}_k$ until convergence.
		\end{itemize}
		\item[] \textbf{CM-step 2}: Given $\boldsymbol{\beta}^{(r)}_k$ and $\sigma^{2(r)}_k$,  updating $\eta_k$ by maximizing \eqref{eta_obj1} with constraint $\eta_k>1$, for $k=1,2,\dots,K$. This can be performed using the \texttt{optimize} function in the \texttt{R} programming language. \\ 
		\item [] Repeat the above E- and CM-steps until convergence.
	\end{enumerate}
\end{algorithm}

\subsection{Semi-parametric mixtures of contaminated Gaussian linear experts}
Although the CG-MoLE model \eqref{CG-MoLE} provides a modelling framework that is robust in the presence of mild outliers, its applicability is limited by the assumption that the mixing proportion functions $\pi_k(\mathbf{t}|\boldsymbol{\gamma}_k)$ are multinomial logistic functions. To enhance the capability of the CG-MoLE model, we propose modeling the mixing proportion functions non-parametrically. In this section, we extend the CG-MoLE model \eqref{CG-MoLE} by introducing the semi-parametric CG-MoLE (S-CG-MoLE) model which assumes that the mixing proportion functions are non-parametric functions of the covariates $\mathbf{t}$ whereas the other model parameters are as defined in the CG-MoLE model. For estimating the non-parametric mixing proportions, we develop a local-likelihood kernel-based estimator obtained using the ECM algorithm. Kernel estimators are known to be subject to the curse-of-dimensionality (see \cite{nagler2016}). To avoid this problem, we only consider a single covariate $t$. Note, however, that the proposed methods can be extended to the case of a multivariate $\mathbf{t}$. Directions towards this end are provided in future studies. 
\subsubsection{Model definition}
Let $Z$ be a component indicator variable with a discrete distribution $P(Z=k|T=t)=\pi_k(t)$, for $k=1,2,\dots,K$, where each $\pi_k(t)>0$ and $\sum_{i=1}^K\pi_k(t)=1$, for all $t$. Given that $Z=k$, $\mathbf{X}=\mathbf{x}$ and $T=t$, $y$ is assumed to follow a contaminated Gaussian distribution with density function $\mathcal{CN}\{y|\mathbf{x};\boldsymbol{\theta}_k\}$ \eqref{CG_pdf}. Given only $\mathbf{X}=\mathbf{x}$ and $T=t$, $y$ follows a mixture of contaminated Gaussian distributions with density function
\begin{eqnarray}\label{S-CG-MoLE}
	f(y|\mathbf{X}=\mathbf{x},T=t)=\sum^{K}_{k=1}\pi_k(t)\mathcal{CN}\{y|\mathbf{x};\boldsymbol{\theta}_k\},
\end{eqnarray}
where $\pi_k(t)$ is assumed to be a smooth unknown function of the covariate $t$. Model \eqref{S-CG-MoLE} is the proposed semi-parametric CG-MoLE (S-CG-MoLE) model.\\
The estimation of model \eqref{S-CG-MoLE} is similar to the estimation of model \eqref{CG-MoLE}; the only difference is that the estimation of the mixing proportion functions is performed non-parametrically using a kernel-based estimator instead of parametrically using the Newton-Raphson algorithm. We turn to the estimation of model \eqref{S-CG-MoLE} in the next section.
\subsubsection{Model estimation}
Consider a random sample $\{(\mathbf{x}_i,t_i,y_i):i=1,2,\dots,n\}$ form the S-CG-MoLE model. The corresponding observed log-likelihood function is 
\begin{eqnarray}
	\ell\{\boldsymbol{\Psi}\}=\sum_{i=1}^n\text{ln}\sum_{k=1}^K\pi_k(t_i)[\alpha_k\mathcal{N}\{y_i|\mathbf{x}_i^\top\boldsymbol{\beta}_k,\sigma^2_k\}+(1-\alpha_k)\mathcal{N}\{y_i|\mathbf{x}_i^\top\boldsymbol{\beta}_k,\eta_k\sigma^2_k\}],
\end{eqnarray}
where $\boldsymbol{\Psi}=(\boldsymbol{\pi},\boldsymbol{\theta})$, $\boldsymbol{\pi}=(\boldsymbol{\pi}_1(t),\boldsymbol{\pi}_2(t),\dots,\boldsymbol{\pi}_K(t))$, with $\boldsymbol{\pi}_k=\{\pi_k(t_i):i=1,2,\dots,n\}$, and $\boldsymbol{\theta}=(\boldsymbol{\theta}_1,\boldsymbol{\theta}_2,\dots,\boldsymbol{\theta}_K)$, with $\boldsymbol{\theta}_k=(\alpha_k,\boldsymbol{\beta}_k,\sigma^2_k,\eta_k)$.\\
As with the CG-MoLE model \eqref{CG-MoLE}, the estimation of the S-CG-MoLE model is performed by maximizing the observed data log-likelihood $\ell\{\boldsymbol{\Psi}\}$ using the ECM algorithm. In the case of the S-CG-MoLE model, the two CM-steps arise from the partition $\boldsymbol{\Psi}=\{\boldsymbol{\Psi}_1,\boldsymbol{\Psi}_2\}$, where $\boldsymbol{\Psi}_1=\{\alpha_k,\boldsymbol{\beta}_k,\sigma^2_k,\boldsymbol{\pi}_k\}_{k=1}^K$ and $\boldsymbol{\Psi}_2=\{\eta_k\}_{k=1}^K$. The corresponding complete-data log-likelihood is 
\begin{eqnarray}\label{llc2}
	\ell_c\{\boldsymbol{\Psi}\}=\ell_{1c}\{\boldsymbol{\pi}\}+\ell_{2c}\{\boldsymbol{\alpha}\}+\ell_{3c}\{\boldsymbol{\vartheta}\},
\end{eqnarray}
where $\ell_{1c}\{\boldsymbol{\pi}\}=\sum_{i=1}^n\sum_{k=1}^Kz_{ik}\text{ln}\pi_k(t_i)$. The complete-data log-likelihoods $\ell_{2c}\{\boldsymbol{\alpha}\}$ and $\ell_{3c}\{\boldsymbol{\vartheta}\}$ are as defined in \eqref{llc1}.
\paragraph{E-step}
In the E-step, at the $r^{th}$ iteration of the ECM algorithm, we calculate the posterior probability that the $i^{th}$ data point belongs to the $k^{th}$ component, denoted $z^{(r)}_{ik}$, and the posterior probability that the $i^{th}$ data point is not an outlier in component $k$, denoted $v^{(r)}_{ik}$, respectively, as
\begin{eqnarray}
	z_{ik}^{(r)}&=&\frac{\pi^{(r-1)}_{k}(t_i)\mathcal{CN}\{y_i|\mathbf{x}_i;\boldsymbol{\theta}^{(r-1)}_k\}}{\sum_{\ell=1}^K\pi^{(r-1)}_{\ell}(t_i)\mathcal{CN}\{y_i|\mathbf{x}_i;\boldsymbol{\theta}^{(r-1)}_\ell\}}\label{resp3},\\
	v^{(r)}_{ik}&=&\frac{\alpha^{(r-1)}_k\mathcal{N}\{y_i|\mathbf{x}^\top_i\boldsymbol{\beta}^{(r-1)}_k,\sigma^{2(r-1)}_k\}}{\mathcal{CN}\{y_i|\mathbf{x}_i;\boldsymbol{\theta}^{(r-1)}_k\}}.\label{resp4}
\end{eqnarray}
Substituting $z_{ik}$ with $z_{ik}^{(r)}$ and $v_{ik}$ with $v^{(r)}_{ik}$ in \eqref{llc2}, we obtain
\begin{eqnarray}
	Q\{\boldsymbol{\Psi}|\boldsymbol{\Psi}^{(r-1)}\}=Q\{\boldsymbol{\pi}|\boldsymbol{\pi}^{(r-1)}\}+Q\{\boldsymbol{\alpha}|\boldsymbol{\alpha}^{(r-1)}\}+Q\{\boldsymbol{\vartheta}|\boldsymbol{\vartheta}^{(r-1)}\},
\end{eqnarray}
where 
\begin{eqnarray}
	Q\{\boldsymbol{\pi}|\boldsymbol{\pi}^{(r-1)}\}&=&\sum_{i=1}^n\sum_{k=1}^Kz^{(r)}_{ik}\text{ln}\pi_k(t_i).\nonumber
\end{eqnarray}
The functions $Q\{\boldsymbol{\alpha}|\boldsymbol{\alpha}^{(r-1)}\}$ and $Q\{\boldsymbol{\vartheta}|\boldsymbol{\vartheta}^{(r-1)}\}$ are defined similar to \eqref{expllc1}.

\paragraph{CM-step 1}
In the first CM-step, at the $r^{th}$ iteration, of the ECM algorithm, we calculate $\boldsymbol{\Psi}^{(r)}_1$, by maximizing $Q\{\boldsymbol{\Psi}|\boldsymbol{\Psi}^{(r-1)}\}$ with $\boldsymbol{\Psi}_2$ fixed at $\boldsymbol{\Psi}^{(r-1)}_2$. In particular, by maximizing $Q\{\boldsymbol{\alpha}|\boldsymbol{\alpha}^{(r-1)}\}$ with respect to $\alpha_k$, for $k=1,2,\dots,K$, and $Q\{\boldsymbol{\vartheta}|\boldsymbol{\vartheta}^{(r-1)}\}$ with respect to $\boldsymbol{\beta}_k$ and $\sigma^2_k$, for $k=1,2,\dots,K$, respectively, we obtain
\begin{eqnarray}
	\alpha^{(r)}_k&=&\frac{1}{n^{(r)}_k}\sum_{i=1}^nz^{(r)}_{ik}v^{(r)}_{ik}\label{alpha_est2},\\
	\boldsymbol{\beta}^{(r)}_k&=&(\mathbf{X}^\top \mathbf{W}^{(r)}_k\mathbf{X})^{-1}\mathbf{X}^\top \mathbf{W}^{(r)}_k\mathbf{y}\label{beta_est2},\\
	\sigma^{2(r)}_k&=&\frac{1}{n^{(r)}_k}\sum_{i=1}^n w^{(r)}_{ik}(y_i-\mathbf{x}^\top_i\boldsymbol{\beta}^{(r)}_k)^2.\label{sigma_est2}
\end{eqnarray}
Equations \eqref{alpha_est2} - \eqref{sigma_est2} are defined similarly to equations \eqref{alpha_est}-\eqref{sigma_est}.\\
To calculate the non-parametric estimate of $\pi_k(t)$, for $k=1,2,\dots,K$, we will make use of a kernel-based local-linear estimator \cite{fan1996}. We chose the local-linear estimator because of its good properties discussed in detail by \cite{fan1992}. We first note that the mixing proportion function can be defined as 
\begin{eqnarray}
	\pi_k(t_i)=\mathbb{E}\{Z_{ik}|T_i=t_i\}
\end{eqnarray}
Thus, $\pi_k(t_i)$ is a regression function with a response variable $Z_{ik}$ and covariate $t_i$. Since we do not observe the component indicator $Z_{ik}$, we make use of $z^{(r)}_{ik}$, its estimate obtained at the E-step of the ECM algorithm, as the response variable.\\
For $t$ in a neighbourhood of a local point $u\in \mathcal{U}$, where $\mathcal{U}$ is the set of local points on the domain of the covariate $t$, $\pi_k(t)$ can be locally approximated by a linear function
\begin{eqnarray}
	\pi_k(t)\approx \pi_k(u)+\pi^\prime_k(u)[t-u]\equiv a_k+b_k(t-u).
\end{eqnarray}
This gives rise to the following locally weighted least squares criterion
\begin{eqnarray}\label{criterion1}
	-\frac{1}{2}\sum_{i=1}^n\bigg[z^{(r)}_{ik}-a_k-b_k(t_i-u)\bigg]^2\mathcal{K}_h(t_i-u),
\end{eqnarray}
where $\mathcal{K}_h(\cdot)=\mathcal{K}(\cdot/h)/h$ is a rescaled kernel function that assigns weights to points in the neighbourhood of $u$ and $h$ is a smoothing parameter that controls the size of the local neighbourhood around $u$. The constant $1/2$ is included for computational convenience.\\
Maximizing \eqref{criterion1} with respect to $a_k$ and $b_k$, we obtain the local-linear estimator 
\begin{eqnarray}\label{llest_pi}
	\pi^{(r)}_k(u)\equiv a^{(r)}_k=\frac{\sum_{i=1}^n[s_2(u)-s_1(u)(t_i-u)]\mathcal{K}_h(t_i-u)z^{(r)}_{ik}}{s_2(u)s_0(u)-s^2_1(u)},
\end{eqnarray}
where $s_j(u)=\sum^{n}_{i=1}\mathcal{K}_h(t_i-u)(t_i-u)^j$, for $j=0,1,2$.\\
The non-parametric curve estimate is obtained by using \eqref{llest_pi} over all $u\in \mathcal{U}$. To obtain $\pi^{(r)}_k(t_i)$, for $t_i\notin \mathcal{U}$, we make use of linear interpolation. Note that, if the sample size is not large, one can utilize the observed values of the covariate $t$ as local points.
\paragraph{CM-step 2}
In the second CM-step, at the $r^{th}$ iteration of the ECM algorithm, we calculate $\boldsymbol{\Psi}^{(r)}_2$ by maximizing $Q\{\boldsymbol{\Psi}|\boldsymbol{\Psi}^{(r-1)}\}$ with $\boldsymbol{\Psi}_1$ fixed at $\boldsymbol{\Psi}^{(r)}_1$ obtained at CM-step 1. In particular, we maximize 
\begin{eqnarray}\label{eta_obj1}
	-\frac{1}{2}\sum_{i=1}^n z_{ik}^{(r)}(1-v^{(r)}_{ik})\text{ln}\eta_k-\frac{1}{2}\sum_{i=1}^nz^{(r)}_{ik}\frac{(1-v^{(r)}_{ik})}{\eta_k}\frac{(y_i-\mathbf{x}^\top_i\boldsymbol{\beta}^{(r)}_k)^2}{\sigma^{2(r)}_k},
\end{eqnarray}
with respect to $\eta_k$, given the constraint $\eta_k>1$, for $k=1,2,\dots,K$.\\
The above ECM algorithm is summarized in Algorithm \ref{Alg2}.
\begin{algorithm}
	\caption{ECM Algorithm for fitting S-CG-MoLE}
	\label{Alg2}
	Let $r$ represent the index of the current iteration, given the estimates from the previous iteration $\boldsymbol{\theta}^{(r-1)}_k=(\boldsymbol{\beta}^{(r-1)}_{k},\sigma^{2(r-1)}_k,\alpha^{(r-1)}_k,\eta^{(r-1)}_k)$ and $\boldsymbol{\psi}^{(r-1)}_k=( \boldsymbol{\gamma}^{(r-1)}_k,\boldsymbol{\beta}^{(r-1)}_k,\sigma^{2(r-1)}_k,\alpha^{(r-1)}_k,\eta^{(r-1)}_k\ )$, the algorithm alternates between the E-step and two CM-steps until convergence.
	\begin{enumerate}
		\item[] \textbf{E-step}: Computing the posterior probabilities $z^{(r)}_{ik}$ and $v^{(r)}_{ik}$, for $i=1,2,\dots,n$ and $k=1,2,\dots,K$, using \eqref{resp3} and \eqref{resp4}, respectively.\\
		
		\item[] \textbf{CM-step 1}: Updating the parameters $\alpha_k$, $\boldsymbol{\beta}_k$, $\sigma^2_k$, for $k=1,2,\dots,K$, and the non-parametric mixing proportion functions $\pi_k(t_i)$, for $i=1,2,\dots,n$ and $k=1,2,\dots,K-1$. 
		\begin{itemize}
			\item \textbf{Parametric estimation:} Given $\eta^{(r-1)}_k$, calculating $\alpha^{(r)}_k$, $\boldsymbol{\beta}^{(r)}_k$ and $\sigma^{2(r)}_k$ using \eqref{alpha_est}, \eqref{beta_est} and \eqref{sigma_est}
			\item \textbf{Non-parametric estimation:} For each local point $u\in \mathcal{U}$ in the domain of the covariate $t$, calculate $\pi^{(r)}_k(u)$ using \eqref{llest_pi} and obtain $\pi^{(r)}_k(t_i)$, for $t_i\notin \mathcal{U}$, using linear interpolation. 
		\end{itemize}
		\item[] \textbf{CM-step 2}: Given $\boldsymbol{\beta}^{(r)}_k$ and $\sigma^{2(r)}_k$,  update $\eta_k$ by maximizing \eqref{eta_obj1} with constraint $\eta_k>1$, for $k=1,2,\dots,K$. This can be performed using the \texttt{optimize} function in the \texttt{R} programming language. \\ 
		\item [] Repeat the above E- and CM-steps until convergence.
	\end{enumerate}
\end{algorithm}
\newpage
\subsection{Model selection}
For the proposed S-CG-MoLE \eqref{S-CG-MoLE}, model selection involves choosing $K$, the number of experts, $h$, the smoothing parameter, and $\mathcal{K}(\cdot)$, the kernel function. To select $K$, we use the Bayesian information criterion (BIC) defined as
\begin{eqnarray}
	\text{BIC}(K)=-2\hat{\ell}+df\times \text{log}(n),
\end{eqnarray}
where $\hat{\ell}$ is the maximum value of the log-likelihood for the fitted model and $df$ is the degrees of freedom measured by the number of parameters in the fitted model. Note that the proposed S-CG-MoLE model has both a parametric and non-parametric part. Thus, we define $df=df_1+df_2$, where $df_1=3K+K(p+1)$ and $df_2=K\times \text{EDF}$ are the degrees of freedom for the parametric and non-parametric part of the model, respectively. EDF is the effective degrees of freedom of a one-dimensional smooth unknown function such as $\pi_k(\cdot)$ \cite{green1994}. Following the popular approach in the literature (\cite{huang2013}, \cite{xiang2020}, \cite{skhosana2024}), we define the EDF as
\begin{eqnarray}
	\text{EDF}=\tau_Kh^{-1}|\mathcal{T}|\{\mathcal{K}(0)-\frac{1}{2}\int \mathcal{K}^2(u)du\},
\end{eqnarray}
where $\mathcal{T}$ is the support of $t$ and 
\begin{eqnarray}
	\tau_K=\frac{\mathcal{K}(0)-\frac{1}{2}\int \mathcal{K}^2(u)du}{\int \{\mathcal{K}(u)-\frac{1}{2}\mathcal{K}*\mathcal{K}(u)\}^2du}.
\end{eqnarray}
The bandwidth $h$ is chosen using the same cross-validation procedure in \cite{huang2012}.\\
Finally, the choice of the kernel function is not usually that important because it does not have any influence on the convergence of the local-linear estimator \cite{wu2006}. For non-parametric mixtures of regressions, in particular, the Gaussian kernel
\begin{eqnarray}
	\mathcal{K}(u)=\big(\sqrt{2\pi}\big)^{-1}\text{exp}(-u^2/2)
\end{eqnarray}
is, by far, one of the most widely used kernel functions in practice \cite{xiang2018}.
\subsection{Model-based clustering}
For any contaminated Gaussian mixture model, an observation can be classified in two stages. In the first stage, we determine its cluster membership. Thereafter, in the second stage, we determine whether or not the observation is an outlier in that cluster. For the S-CG-MoLE model \eqref{S-CG-MoLE}, the first stage is performed using the maximum a posteriori (MAP) operator
\begin{equation}\label{MAP}
	\text{MAP}(\hat{z}_{ik})=\begin{cases}
		1\quad\text{if $\max_j\hat{z}_{ij}$ occurs in component $k=j$}\\
		0\quad\text{otherwise},
	\end{cases}
\end{equation}
where $\hat{z}_{ik}$ is the value of $z^{(r)}_{ik}$ at convergence of the ECM algorithm \ref{Alg2}. Using \eqref{MAP}, the $i^{th}$ observation $(\mathbf{x}_i,t_i,y_i)$ will be assigned to cluster $j$. This is followed by the second stage in which the observation $(\mathbf{x}_i,t_i,y_i)$ assigned to cluster $j$ will be labelled as an outlier if $\hat{v}^{(r)}_{ik}<0.5$, where $\hat{v}_{ik}$ is the value of $v^{(r)}_{ik}$ at convergence of the ECM algorithm \ref{Alg2}.
\section{Simulation Study}\label{sec4}
In this section, we perform an extensive numerical study to evaluate the performance of the proposed methods. The proposed S-CGMoLE model is compared with the 
\begin{enumerate}
	\item Gaussian mixture of linear regressions (GMLRs) (\ref{model1});
	\item contaminated Gaussian mixture of linear regressions (CGMLRs) \cite{mazza2020};
	\item Gaussian mixture of linear experts (GMoLE) (\ref{model2});
	\item contaminated Gaussian mixture of linear experts (CGMoLE) (\ref{CG-MoLE});
	\item semi-parametric Gaussian mixture of linear experts (S-GMoLE) with non-parametric gating functions (\ref{model3}).
\end{enumerate}
The simulations are conducted on the $\texttt{R}$ programming language \cite{R2023}. The GMLRs and GMoLE are estimated using the $\texttt{regmixEM}$ function and $\texttt{hmeEM}$ function, respectively, from the $\texttt{R}$ package $\texttt{mixtools}$ \cite{benaglia2010}. There are no existing packages for fitting CGMLRs, CGMoLE and the S-GMoLE. Therefore, maximum likelihood via the EM and/or ECM estimation procedures were written in $\texttt{R}$ for the purpose of estimating these models. To evaluate the estimation accuracy of each method, we make use of the following measures:
\begin{eqnarray}
	\text{MSE}(\hat{\boldsymbol{\pi}}(x))&=&\frac{1}{n}\sum_{i=1}^n\sum_{k=1}^K\big(\pi_k(x_i)-\hat{\pi}_k(x_i)\big)^2,\nonumber
\end{eqnarray}
where $\hat{\pi}_k(x)$ and $\pi_k(x)$ are the estimated and true values of the mixing proportion function $\pi_k(x)$, respectively. 
\begin{eqnarray}
	\text{MSE}(\hat{\theta}_k)&=&(\hat{\theta}_k-\theta_k)^2\,\,\text{and }\, \text{BIAS}(\hat{\theta}_k)=\hat{\theta}_k-\theta_k,\nonumber
\end{eqnarray}
where $\hat{\theta}_k$ and $\theta_k$ are the estimated and true values of the parameter $\theta_k$, respectively. 
\subsection{Numerical studies}
The data for this study is generated from a $K=2$ component semi-parametric mixture of linear regressions. Each $y_i$, for $i=1,2,\dots,n$, is independently generated as follows
\begin{eqnarray}
	y_i=\begin{cases}
		x_i+\epsilon_{i1}\quad \text{if } z_i=1\\
		4+x_i+\epsilon_{i2}\quad \text{if } z_i=2,
	\end{cases}\nonumber
\end{eqnarray}
where $z_i$ is the component indicator of the $i^{th}$ data point, with distribution $P(z_i=1)=\pi_1(x_i)=0.1+0.8\text{sin}(\pi x_i)$, $x_i$ is generated from a uniform distribution on the interval $(0,1)$ and $\epsilon_{ik}$ is the error term of the $k^{th}$ linear regression model. We assume that the error terms $\epsilon_{i1}$ and $\epsilon_{i2}$ follow the same distribution as $\epsilon$. We consider the following distributions for $\epsilon$:
\begin{enumerate}
	\item $\epsilon\sim\mathcal{N}\{0,1\}$;
	\item $\epsilon\sim 0.95\mathcal{N}\{0,\sigma^2\}+0.05\mathcal{N}\{0,\eta\sigma^2\}$, with $\sigma=1$ and $\eta=20$;
	\item $\epsilon\sim \text{t-distribution}(\nu)$, with $\nu=3$ degrees of freedom;
	\item $\epsilon\sim\mathcal{N}\{0,1\}$, with $10\%$ of the $y$-values randomly substituted by noise data points generated from the uniform distribution on the interval $(-15,15)$.
\end{enumerate}
The first scenario is used to evaluate the performance of the proposed methods when the data is not contaminated by outliers. The second and third scenarios are used to evaluate the performance of the proposed methods when the data has a heavy-tailed distribution due to the presence of outliers in the data. The second scenario also serves to evaluate the performance of the proposed Algorithm \ref{Alg2} for fitting the S-CG-MoLE model. 
In each scenario, we generate $100$ samples of sizes $n=200$, $500$ and $1000$. Tables \ref{res1}-\ref{res3} give the MSE and BIAS of the parameter estimates for $n=200$, $500$, and $1000$, respectively. In addition, Table \ref{res4} gives the average and standard deviation of the MSEs of the fitted mixing proportion functions and parameters.\\
The main takeaways from Tables \ref{res1}-\ref{res4} are as follows:
\begin{itemize}
	\item For scenario (a), the models SGMoE, GMoE, and GMLRs were the overall best models as expected since the component error distribution is Gaussian. However, the model's performance is virtually the same as the performance of the proposed SCGMoE model for large $n=1000$. This shows that the proposed SCGMoE model performs well even when the data does not have outliers.
	\item For scenario (b), the proposed SCGMoE, CGMoE, and CGMLRs were the overall best models as expected since the component error distribution follows a contaminated Gaussian. On the other hand, the performance of the models SGMoE, GMoE, and GMLRs does not improve even as we increase the sample size to $n=1000$. This shows the negative effect of outliers on the parameter estimates when the component distributions are assumed to be Gaussian. 
	\item For scenario (c), the models SCGMoE, SGMoE, CGMoE, and CGMLRs had the overall best performance, and the performance is similar across the models. On the other hand, the models GMoE and GMLRs performed the worst, and the performance does not improve as we increase the sample size $n=1000$. Notice that the SGMoE model performed well even if it assumed that the component error distribution is Gaussian. This highlights the importance of correct model specification for the mixing proportions in the presence of outliers. In the event of misspecification, the resulting bias leads to biases in the component-specific parameter estimates, see the results for the GMLRs and GMoE, where the functional form of the mixing proportions is misspecified.
	\item For scenario (d), the models CGMLRs, CGMoE and the proposed S-CGMoE are the overall best models. On the other hand, the performance of the models GMLRs, GMoE, and SGMoE does not improve even when we increase the sample size to $n=1000$. 
\end{itemize}
In summary, this simulation study showed the importance of using a heavy-tailed component error distribution when the data is heavy-tailed (scenarios (b) and (c)) and also when the data is contaminated by noisy data points (scenario (d)). Moreover, it also showed the effect of mis-specifying the mixing proportion functions on the model parameter estimates when the data has outliers and the component error distribution is assumed to be Gaussian. 

\begin{table}[!htbp]
\scriptsize
	\centering
	\caption{\scriptsize MSE (BIAS) of the fitted model parameters for $n=200$ (The values are multiplied by 100)}\label{res1}
	\begin{tabular}{C{1.5cm}C{1.5cm}C{1.5cm}C{1.5cm}C{1.5cm}C{1.5cm}C{1.5cm}}
		\hline
		&\text{GMLRs}&\text{CGMLRs}&\text{GMoE}&\text{CG-MoE}&\text{S-G-MoE}&\text{S-CG-MoE}\\
		\hline
		&\multicolumn{6}{c}{Scenario (a)}\\
		\hline
		$\beta_{10}$&0.010 &0.001 &0.001 &0.001 &0.003 &0.030 \\
		&(-0.77)&(-0.33)&(0.053)&(-0.10)&(-0.51)&(-1.652)\\
		$\beta_{11}$&0.080 &0.090&0.021 &0.026 &0.020 &0.034 \\
		& (2.78)& (3.05)& (1.445)& (1.61)& (1.40)& (1.86)\\
		$\beta_{20}$&0.022 &0.012 &0.060 &0.001 &0.004 &0.021\\
		& (1.476)& (1.08)& (2.353)& (0.18)& (0.66)& (-1.45)\\
		$\beta_{21}$&0.002 &0.002 &0.001 &0.001 &0.001&0.001\\
		& (0.45)& (0.487)& (-0.200)& (0.12)& (-0.19)& (0.29)\\
		$\sigma_{1}$&0.004 &0.500 &0.002 &0.781 &0.010&0.770 \\
		&(-0.63)& (-7.08)& (-0.403)& (-8.84)& (-0.90)&(-8.75)\\
		$\sigma_{2}$&0.100 &0.970 &0.120 &0.930 &0.050 &0.730 \\
		& (-3.18)& (-9.83)& (-3.48)& (-9.63)& (-2.26)& (-8.56)\\
		\hline
		&\multicolumn{6}{c}{Scenario (b)}\\
		\hline
		$\beta_{10}$&9.41 &0.07&97.51&0.020&0.520 &0.000 \\
		&(-30.68)&(2.69)& (-98.75)& (1.560)& (7.240)& (-0.560)\\
		$\beta_{11}$&90.96 &0.53 &422.38 &0.420 &1.23 &0.35 \\
		& (95.37)& (-7.28)& (205.52)& (-6.470)& (-11.08)& (-5.91)\\
		$\beta_{20}$&0.11 &0.02 &1.11 &0.00 &0.42 &0.03 \\
		& (3.31)& (1.42)& (-10.53)& (0.06)& (-6.51)& (-1.75)\\
		$\beta_{21}$&672.26 &0.01 &449.02 &0.00 &1.11 &0.00 \\
		& (-259.28)& (-0.74)& (-211.90)& (-0.33)& (-10.52)&(-0.32)\\
		$\sigma_{1}$&48.24 &0.12 &52.10 &0.14 &28.88 &0.25 \\
		& (69.46)& (-3.47)& (72.18)& (-3.74)& (53.74)& (-5.00)\\
		$\sigma_{2}$&114.16 &0.27 &72.07 &0.24 &45.90 &0.150 \\
		&(106.85)& (-5.20)& (84.89)& (-4.900)& (67.75)&(-3.820)\\
		\hline
		&\multicolumn{6}{c}{Scenario (c)}\\
		\hline
		$\beta_{10}$&1.764 &0.062 &206.998 &0.571 &1.046 &1.621 \\
		& (-13.281)& (-2.484)& (-143.874)& (-7.556)& (-10.226)& (-12.733)\\
		$\beta_{11}$&10.74 &0.102 &976.696 &1.602 &2.165 &1.878 \\
		&(32.773)& (3.202)& (312.521)& (12.658)&(14.713)& (13.704)\\
		$\beta_{20}$&20.549&0.192 &8.726 &0.007 &0.018 &0.217\\
		& (45.331)& (4.377)& (-29.54)& (0.859)& (1.328)& (-4.655)\\
		$\beta_{21}$&268.752 &0.782 &1325.462 &0.043 &0.555 &0.072 \\
		& (-163.937)& (-8.841)& (-364.069)& (-2.066)& (-7.45)& (-2.686)\\
		\hline
		&\multicolumn{6}{c}{Scenario (d)}\\
		\hline
		$\beta_{10}$&76.277 &0.019 &475.948 &0.001 &4.805&0.017\\
		& (-87.337)& (1.381)& (-218.162)& (-0.200)& (21.92)& (-1.323)\\
		$\beta_{11}$&619.26 &0.007 &1854.437 &0.016&0.047 &0.005 \\
		& (248.849)& (-0.823)& (430.632)& (1.276)&(2.178)& (0.687)\\
		$\beta_{20}$&16.874 &0.044 &9.442 &0.005 &19.202 &0.001 \\
		& (-41.078)&(2.095)& (-30.728)& (0.734)& (-43.82)&(-0.119)\\
		$\beta_{21}$&575.425 &0.185 &1869.823 &0.114 &1.976 &0.171 \\
		& (-239.88)& (-4.303)& (-432.414)& (-3.374)& (-14.058)& (-4.14)\\
		$\sigma_{1}$&665.253 &0.084 &287.047 &0.135 &182.512 &0.21 \\
		& (257.925)& (-2.895)& (169.425)& (-3.678)&(135.097)& (-4.587)\\
		$\sigma_{2}$&325.515 &0.694 &209.874 &0.396 &96.371 &0.297\\
		& (180.42)& (-8.329)& (144.87)& (-6.29)& (98.169)& (-5.453)\\
		\hline
	\end{tabular}
\end{table}

\begin{table}[htbp]
\scriptsize
\centering
	\caption{\scriptsize MSE (BIAS) of the fitted model parameters for $n=500$ (The values are multiplied by 100)}\label{res2}
	\begin{tabular}{C{1.5cm}C{1.5cm}C{1.5cm}C{1.5cm}C{1.5cm}C{1.5cm}C{1.5cm}}
		\hline
		&\text{GMLRs}&\text{CGMLRs}&\text{GMoE}&\text{CG-MoE}&\text{S-G-MoE}&\text{S-CG-MoE}\\
		\hline
		&\multicolumn{6}{c}{Scenario (a)}\\
		\hline
		$\beta_{10}$&0.005 &0.014 &0.022 &0.019&0.010 &0.001 \\
		& (0.727)& (1.161)& (1.488)& (1.389)& (1.005)& (0.293)\\
		$\beta_{11}$&0.001 &0.001 &0.024 &0.016 &0.014 &0.017\\
		& (-0.090)& (-0.133)& (-1.54)& (-1.272)& (-1.171)& (-1.308)\\
		$\beta_{20}$&0.007 &0.004 &0.020 &0.001 &0.004 &0.009 \\
		& (0.864)& (0.67)& (1.41)& (0.393)& (0.644)& (-0.940)\\
		$\beta_{21}$&0.001 &0.002 &0.002 &0.001 &0.001 &0.001 \\
		& (0.331)& (0.41)& (-0.456)& (-0.259)& (-0.245)& (-0.208)\\
		$\sigma_{1}$&0.002 &0.21 &0.002 &0.324 &0.004 &0.334 \\
		&(-0.468)& (-4.59)& (-0.409)& (-5.695)& (-0.649)& (-5.781)\\
		$\sigma_{2}$&0.012 &0.256 &0.014 &0.28 &0.005&0.205 \\
		& (-1.099)& (-5.06)& (-1.18)& (-5.27)& (-0.697)& (-4.53)\\
		\hline
		&\multicolumn{6}{c}{Scenario (b)}\\
		\hline
		$\beta_{10}$&1.07 &0.00 &39.49 &0.08 &0.47 &0.01 \\
		& (-10.34)& (-0.12)& (-62.84)& (2.86)& (6.88)& (0.95)\\
		$\beta_{11}$&128.1 &0.02 &413.07 &0.17 &0.15&0.1 \\
		& (113.18)& (1.27)& (203.24)& (-4.15)& (-3.91)& (-3.13)\\
		$\beta_{20}$&3.74 &0.00 &3.53 &0.08 &1.45 &0.00\\
		& (-19.35)& (0.7)& (-18.78)& (2.75)& (-12.04)& (0.64)\\
		$\beta_{21}$&591.53 &0.00 &631.8 &0.09 &0.43 &0.05 \\
		&(-243.21)& (0.33)& (-251.36)& (-3.07)& (-6.54)& (-2.25)\\
		$\sigma_{1}$&176.21 &0.08 &81.68 &0.07 &28.1 &0.15 \\
		& (132.75)&(-2.75)& (90.38)& (-2.55)& (53.01)& (-3.81)\\
		$\sigma_{2}$&310.49 &0.05 &102.07 &0.05 &73.76 &0.01 \\
		& (176.21)& (-2.25)& (101.03)& (-2.29)& (85.88)& (-0.92)\\
		\hline
		&\multicolumn{6}{c}{Scenario (c)}\\
		\hline
		$\beta_{10}$&42.363 &0.00&510.69 &0.096 &0.00 &0.474 \\
		& (-65.087)& (0.168)& (-225.984)& (-3.101)& (-0.113)&(-6.888)\\
		$\beta_{11}$&290.835 &0.232 &2172.475 &0.28 &0.519 &0.555\\
		& (170.539)& (4.812)& (466.098)& (5.291)& (7.204)& (7.45)\\
		$\beta_{20}$&0.621 &0.087 &0.53 &0.06 &0.747 &0.354 \\
		& (-7.878)& (-2.951)& (-7.282)& (-2.458)& (-8.645)& (-5.949)\\
		$\beta_{21}$&168.954 &0.097 &2307.075 &0.059 &0.001 &0.14 \\
		& (-129.982)& (-3.115)& (-480.32)& (2.422)& (0.312)& (3.744)\\
		\hline
		&\multicolumn{6}{c}{Scenario (d)}\\
		\hline
		$\beta_{10}$&2.644 &0.117 &155.56 &0.027 &11.144 &0.063 \\
		& (-16.262)& (-3.418)& (-124.724)& (-1.644)& (33.383)& (-2.515)\\
		$\beta_{11}$&296.406 &0.506 &1230.275 &0.083 &0.138 &0.088 \\
		& (172.164)& (7.114)& (350.753)& (2.877)& (-3.719)&(2.959)\\
		$\beta_{20}$&70.995 &0.015 &17.198 &0.004 &29.726 &0.024 \\
		& (-84.258)& (-1.228)& (-41.471)& (-0.624)& (-54.522)& (-1.561)\\
		$\beta_{21}$&315.975 &0.029 &2534.418 &0.003 &0.141 &0.002 \\
		& (-177.757)& (1.701)& (-503.43)& (-0.538)& (3.756)& (-0.408)\\
		$\sigma_{1}$&1199.269 &0.03 &362.874 &0.038 &265.506 &0.068 \\
		& (346.305)& (-1.744)& (190.492)& (-1.954)& (162.944)& (-2.614)\\
		$\sigma_{2}$&483.682 &0.245 &293.23 &0.117 &76.835&0.084 \\
		& (219.928)& (-4.949)& (171.24)& (-3.422)& (87.656)& (-2.901)\\
		\hline
	\end{tabular}
\end{table}

\begin{table}[htbp]
\scriptsize
\centering
	\caption{\scriptsize MSE (BIAS) of the fitted model parameters for $n=1000$ (The values are multiplied by 100)}\label{res3}
	\begin{tabular}{ccccccc}
		\hline
		&\text{GMLRs}&\text{CGMLRs}&\text{GMoE}&\text{CG-MoE}&\text{S-G-MoE}&\text{S-CG-MoE}\\
		\hline
		&\multicolumn{6}{c}{Scenario (a)}\\
		\hline
		$\beta_{10}$&0.00 &0.00 &0.001 &0.001 &0.00 &0.004 \\
		& (0.33)& (0.63)& (-0.245)& (0.253)& (0.010)& (-0.657)\\
		$\beta_{11}$&0.004 &0.01 &0.00 &0.001 &0.00 &0.00 \\
		& (-0.653)& (-0.760)& (0.54)& (-0.078)& (0.008)& (-0.006)\\
		$\beta_{20}$&0.006 &0.00 &0.00 &0.001 &0.003 &0.006 \\
		& (0.806)& (0.614)& (0.59)& (0.299)& (0.547)& (-0.755)\\
		$\beta_{21}$&0.023 &0.02 &0.01 &0.01 &0.01 &0.01 \\
		& (-1.502)& (-1.543)& (-0.88)& (-1.165)& (-1.15)& (-1.161)\\
		$\sigma_{1}$&0.001 &0.10 &0.00 &0.12 &0.001 &0.13 \\
		& (-0.177)& (-3.190)& (-0.14)& (-3.420)& (-0.15)& (-3.56)\\
		$\sigma_{2}$&0.00 &0.19 &0.00 &0.220 &0.002&0.17\\
		& (-0.59)& (-4.38)& (-0.64)& (-4.692)& (-0.499)& (-4.13)\\
		\hline
		&\multicolumn{6}{c}{Scenario (b)}\\
		\hline
		$\beta_{10}$&3.73 &0.00 &115.29 &0.00&0.11&0.01 \\
		& (-19.31)& (-0.18)& (-107.37)& (-0.22)& (3.38)& (-0.97)\\
		$\beta_{11}$&123.99 &0.00&650.2 &0.00&0.04 &0.00 \\
		& (111.35)&(-0.44)& (254.99)& (-0.15)& (-1.99)& (-0.65)\\
		$\beta_{20}$&110.84 &0.01 &55.92 &0.01 &4.06 &0.04\\
		& (-105.28)& (-0.85)& (-74.78)& (-0.89)& (-20.15)& (-1.93)\\
		$\beta_{21}$&30.27 &0.08 &572.83 &0.09&0.22 &0.07 \\
		& (-55.02)& (2.76)& (-239.34)& (2.99)& (4.68)& (2.69)\\
		$\sigma_{1}$&232.54 &0.00&57.17 &0.00&28.79 &0.01 \\
		& (152.49)& (-0.45)& (75.61)& (-0.31)& (53.66)& (-0.97)\\
		$\sigma_{2}$&356.11 &0.00&144.62 &0.00&94.44 &0.00\\
		& (188.71)& (-0.12)& (120.26)& (-0.22)& (97.18)& (0.67)\\
		\hline
		&\multicolumn{6}{c}{Scenario (c)}\\
		\hline
		$\beta_{10}$&62.573 &0.001 &334.622 &0.026 &0.007 &0.116 \\
		& (-79.103)& (-0.24)& (-182.927)& (-1.607)& (0.841)& (-3.402)\\
		$\beta_{11}$&669.793 &0.046 &1864.894 &0.01 &0.00 &0.00 \\
		& (258.804& (-2.138)& (431.844)& (1.018)& (0.072)& (0.074)\\
		$\beta_{20}$&15.413&0.003 &35.746 &0.001&0.713 &0.034 \\
		& (39.259)& (0.584)& (-59.788)& (0.295)& (-8.445)& (-1.833)\\
		$\beta_{21}$&1354.902 &0.199 &3457.934 &0.05 &0.002 &0.094 \\
		& (-368.09)& (-4.46)& (-588.042)& (-2.227)& (-0.467)&(-3.062)\\
		\hline
		&\multicolumn{6}{c}{Scenario (d)}\\
		\hline
		$\beta_{10}$&1.096 &0.002 &378.853 &0.00&12.545 &0.00 \\
		&(10.471)& (0.49)& (-194.642)& (0.213)& (35.419)& (-0.22)\\
		$\beta_{11}$&46.717&0.00&2200.496 &0.002 &1.604 &0.002 \\
		& (68.35)& (-0.139)& (469.094)& (-0.404)& (-12.666)& (-0.425)\\
		$\beta_{20}$&219.251 &0.00&43.403 &0.004 &29.719 &0.012 \\
		& (-148.071)&(-0.188)& (-65.881)& (-0.654)& (-54.515)& (-1.1)\\
		$\beta_{21}$&186.311 &0.00&3031.079 &0.001 &0.006 &0.001 \\
		& (-136.496)& (-0.112)& (-550.552)& (-0.292)& (-0.77)& (-0.277)\\
		$\sigma_{1}$&1507.266 &0.023 &355.283 &0.029 &259.187 &0.043\\
		& (388.235)& (-1.526)& (188.49)& (-1.71)& (160.993)& (-2.068)\\
		$\sigma_{2}$&642.017 &0.137 &342.25 &0.064 &95.851&0.05 \\
		& (253.381)& (-3.699)& (185.0)& (-2.535)& (97.904)& (-2.244)\\
		\hline
	\end{tabular}
\end{table}

\begin{table}[!ht]
\footnotesize
\centering
	\caption{\scriptsize Average (standard deviation) MSEs of the fitted mixing proportions (values are multiplied by 100)}\label{res4}
	\begin{tabular}{ccccccc}
		\hline
		$n$&\text{GMLRs}&\text{CGMLRs}&\text{GMoE}&\text{CG-MoE}&\text{S-G-MoE}&\text{S-CG-MoE}\\
		\hline
		&\multicolumn{6}{c}{Scenario (a)}\\
		\hline
		$200$&12.25 (0.35)&12.26 (0.35)&12.43 (0.47)&12.24 (0.26)&0.75 (0.48)&0.95 (0.63)\\
		$500$&12.39 (0.14)&12.39 (0.14)&12.45 (0.18)&12.38 (0.10)&0.35 (0.21)&0.44 (0.26)\\
		$1000$&12.76 (0.06)&12.76 (0.06)&12.80 (0.08)&12.76 (0.04)&0.20 (0.11)&0.25 (0.14)\\
		\hline
		&\multicolumn{6}{c}{Scenario (b)}\\
		\hline
		$200$&24.4 (14.87)&13.15 (0.48)&28.88 (16.23)&13.07 (0.18)&0.91 (0.77)&1.14 (0.74)\\
		$500$&29.5 (20.04)&12.27 (0.18)&30.17 (16.1)&12.16 (0.09)&0.6 (0.66)&0.55 (0.28)\\
		$1000$&34.28 (21.62)&12.85 (0.08)&31.18 (13.19)&12.83 (0.04)&0.48 (0.61)&0.29 (0.17)\\
		\hline
		&\multicolumn{6}{c}{Scenario (c)}\\
		\hline
		$200$&20.64 (14.74)&12.53 (0.72)&27.49 (13.41)&12.04 (0.17)&0.67 (0.54)&1.50 (1.03)\\
		$500$&24.16 (18.28)&12.49 (2.69)&36.74 (12.59)&12.01 (0.08)&0.30 (0.19)&0.65 (0.40)\\
		$1000$&29.79 (23.02)&12.24 (0.15)&38.23 (11.12)&12.18 (0.04)&0.27 (0.29)&0.37 (0.200)\\
		\hline
		&\multicolumn{6}{c}{Scenario (d)}\\
		\hline
		$200$&39.03 (22.98)&13.09 (0.50)&35.14 (17.70)&12.94 (0.22)&1.66 (7.47)&1.10 (0.73)\\
		$500$&43.46 (20.86)&12.02 (0.20)&35.67 (15.27)&11.86 (0.09)&0.72 (3.43)&0.49 (0.31)\\
		$1000$&49.36 (23.80)&12.34 (0.14)&37.10 (15.94)&12.28 (0.06)&0.40 (0.53)&0.27 (0.16)\\
		\hline
	\end{tabular}
\end{table}
\section{Application}\label{sec5} 
\begin{figure}[!ht]
	\centering
	\includegraphics[width=0.8\textwidth]{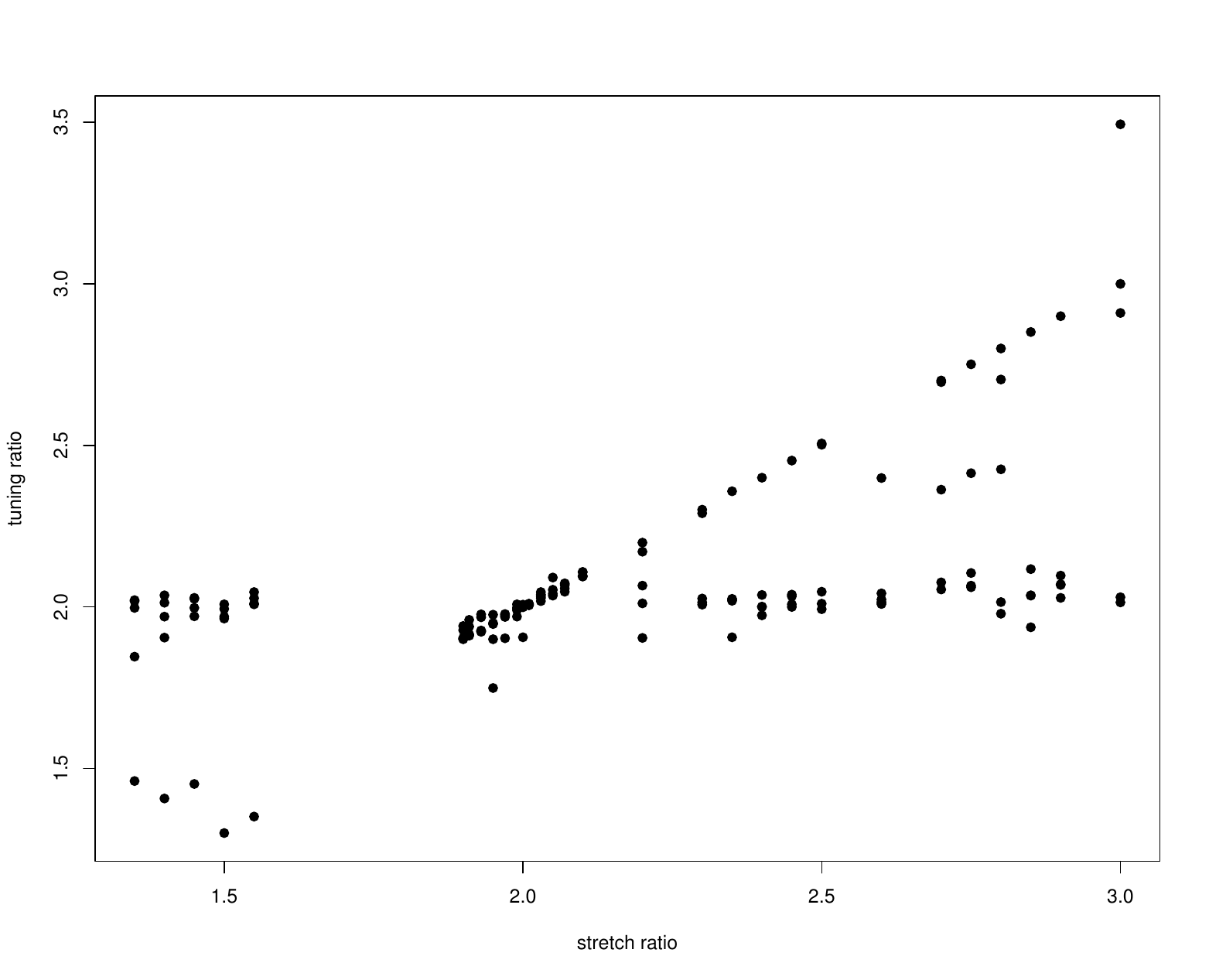}
	\caption{Scatter plot of the tone perception data}
	\label{tonedata}
\end{figure}

In this section, we demonstrate the practical usefulness of the proposed methods on a real dataset on tone perception \cite{cohen1984}. A pure fundamental tone was played to a trained musician, and a stretching ratio was used to add electronically generated overtones. The musician was then asked to tune an adjustable tone to one octave above the fundamental tone. The tuning and stretching ratios were recorded for 150 trials from the same musician. The dataset can be found in the \texttt{R} package \texttt{mixtools} \cite{benaglia2010}. A plot of the dataset is given in \\
 \ref{tonedata}. A mixture analysis of this dataset has been performed by many authors. These studies are in agreement that there are two components in this dataset. Therefore, in the following analysis of this data, we will make use of $K=2$ components.
 
In the first instance, we compare the performance of the proposed S-CG-MoLE model with the following models: GMLRs \eqref{model1}, contaminated GMLRs (CGMLRs) \cite{mazza2020}, GMoLE \cite{jacobs1991}, CG-MoLE \eqref{CG-MoLE}, semi-parametric GMoLE (S-GMoLE) \cite{huang2012}. The results are given in Table \ref{tab:fitted_models1}. The data support the CGMLRs model as the best model according to the BIC ($-424.0539$). This is followed closely by the CG-MoLE ($-419.0688$) and the S-CG-MoLE ($-342.8331$). 
\begin{table}[!ht]
	\caption{Results of the models fitted on the tone perception data.}
	\centering
	\begin{tabular}{|c|c|c|c|c|c|c|}
		\hline
		&\text{GMLRs}&\text{CGMLRs}&\text{GMoE}&\text{CG-MoE}&\text{S-G-MoE}&\text{S-CG-MoE}\\
		\hline
		$\beta_{10}$&-0.0193&0.0034&-0.0295&0.0034&0.0060&0.0033\\ 
		$\beta_{20}$&1.9164&1.9542&1.9132&1.9540&1.9250&1.9603\\
		$\beta_{11}$&0.9923&0.9988&0.9957&0.9988&0.9832&0.9988\\ 
		$\beta_{21}$&0.0426&0.0282&0.0437&0.0283&0.0393&0.0263\\
		$\pi_1$&0.3023&0.4454&-&-&-&\\ 
		$\sigma_1$&0.1328&0.0042&0.1373&0.0042&0.1211&0.0045\\
		$\sigma_2$&0.0462&0.0244&0.0471&0.0244&0.0429&0.0247\\ 
		$\gamma_{10}$&-&-&-0.7921&-0.7409&-&-\\ 
		$\gamma_{11}$&-&-&2.6787&0.1559&-&-\\
		$\eta_1$&-&3506.822&-&3509.089&-&2187.006\\
		$\eta_2$&-&6.6516&-&6.6620&-&7.4175\\
		$\alpha_1$&-&0.7732&-&0.7737&-&0.7303\\
		$\alpha_2$&-&0.5553&-&0.5555&-&0.6678\\
		\hline
		BIC&-247.3224&$\mathbf{-424.0539}$&-245.6109&-419.0688&-105.8673&-342.8331\\
		\hline
	\end{tabular}
	\label{tab:fitted_models1}
\end{table}

\begin{figure}[!ht]
	\centering
	\includegraphics[width=0.75\linewidth]{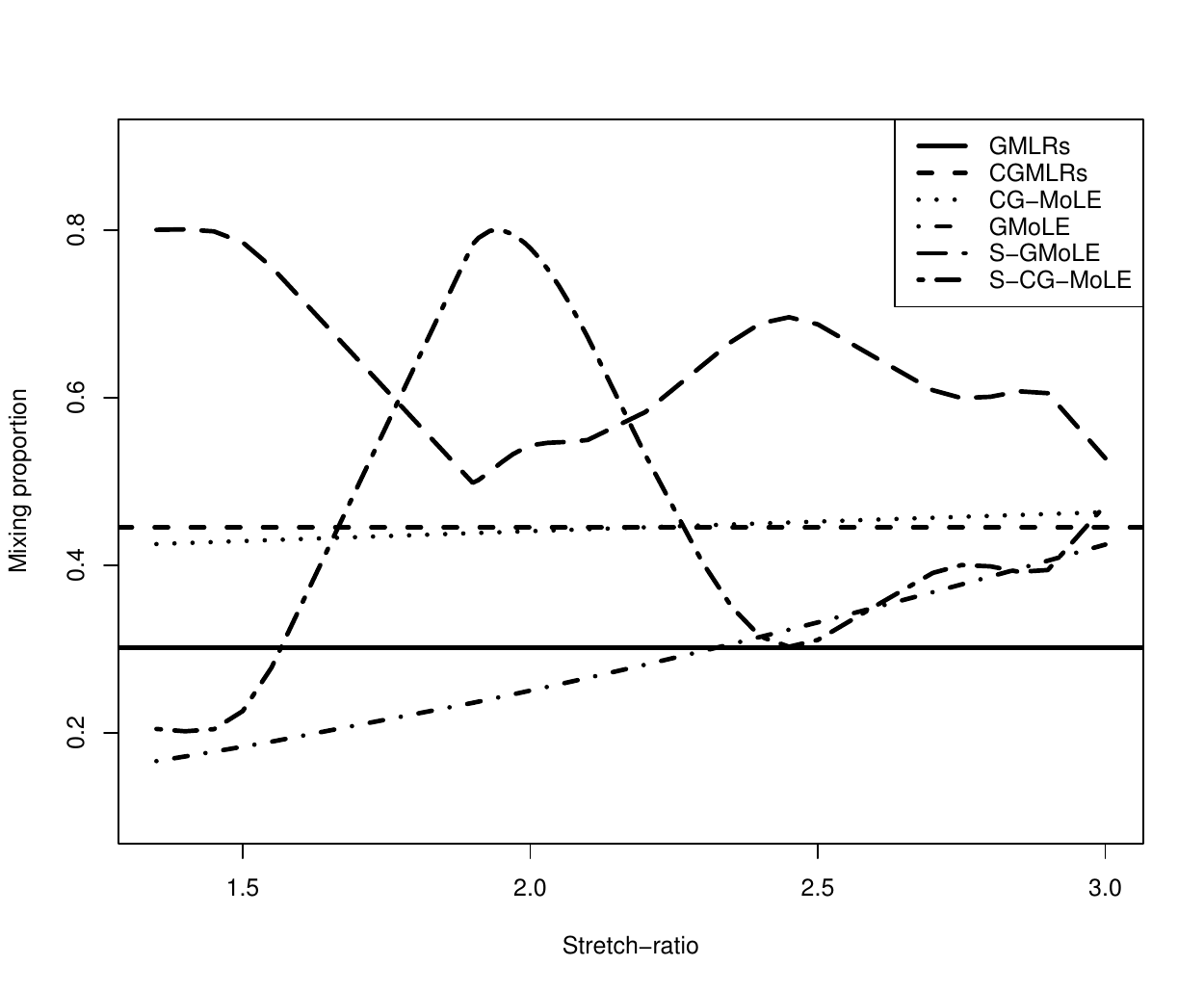}
	\caption{Estimated mixing proportions for the first component of all six fitted models on the tone perception data}
	\label{fitted_mix1}
\end{figure}

\begin{figure}[!ht]
	\centering
	\begin{subfigure}{0.47\textwidth}
		\centering
		\includegraphics[width=\textwidth]{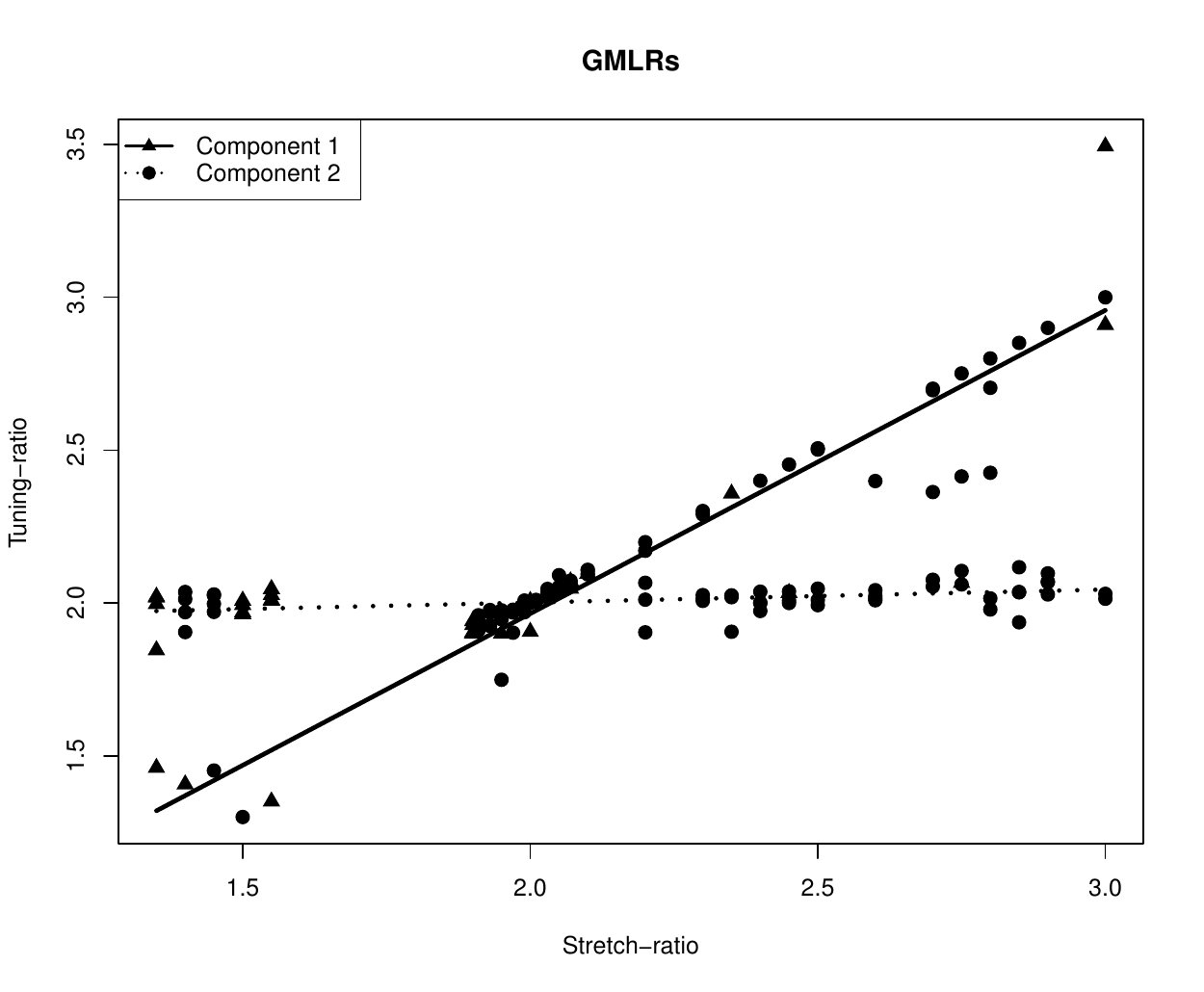}
		\caption{}
		\label{fig:first}
	\end{subfigure}
	\hfill
	\begin{subfigure}{0.47\textwidth}
		\centering
		\includegraphics[width=\textwidth]{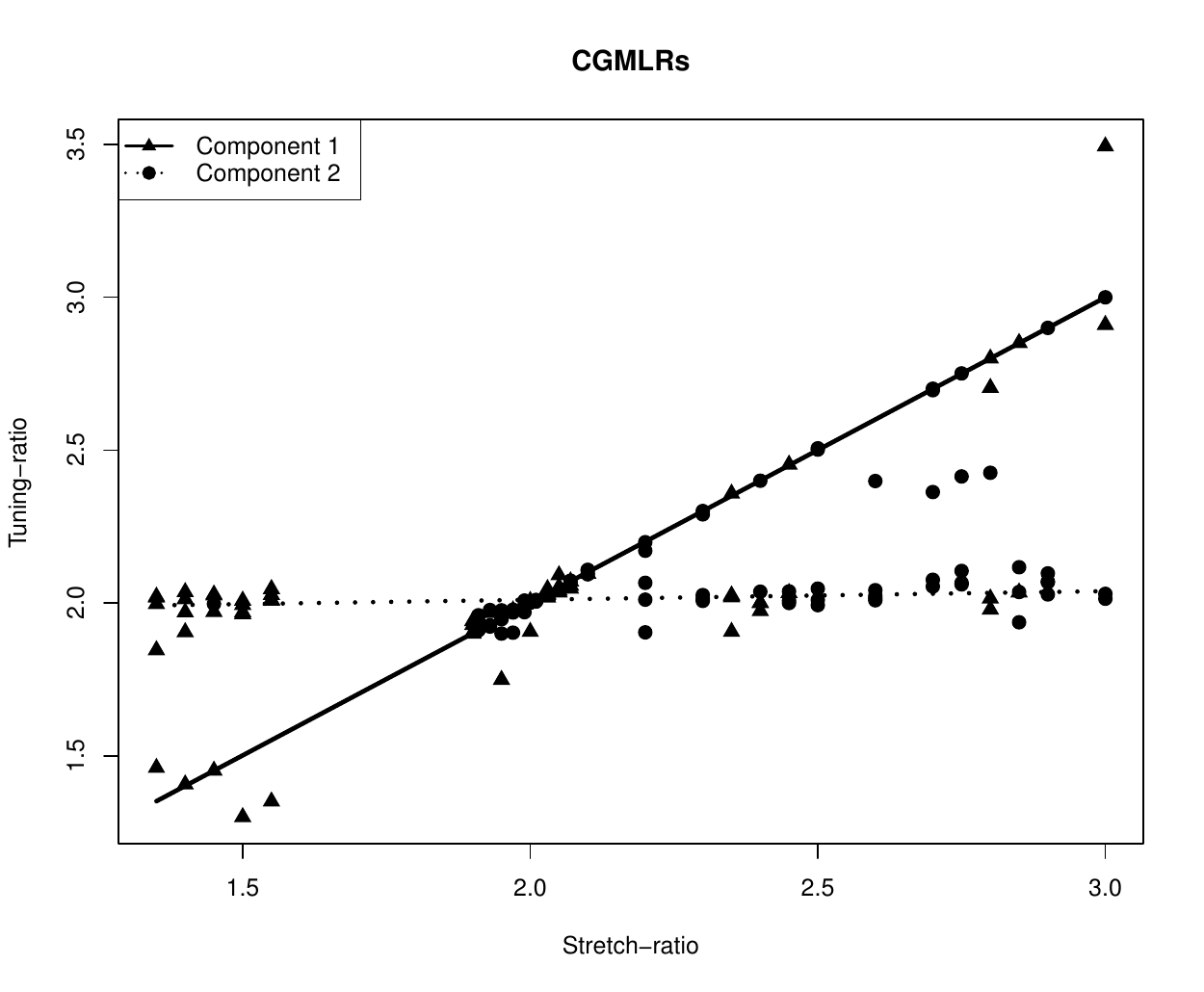}
		\caption{}
		\label{fig:second}
	\end{subfigure}
	\\
	\begin{subfigure}{0.47\textwidth}
		\centering
		\includegraphics[width=\textwidth]{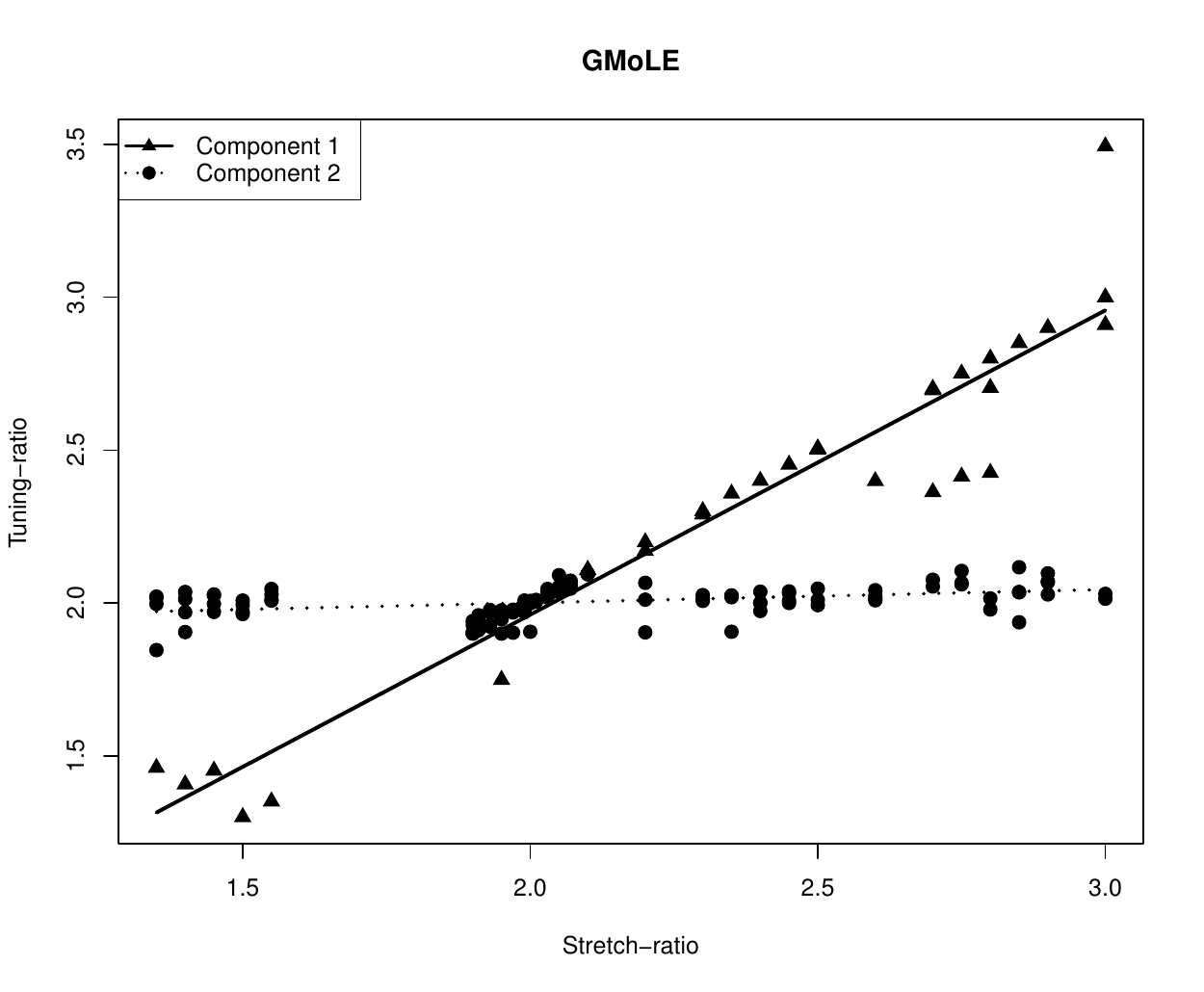}
		\caption{}
		\label{fig:first}
	\end{subfigure}
	\hfill
	\begin{subfigure}{0.47\textwidth}
		\centering
		\includegraphics[width=\textwidth]{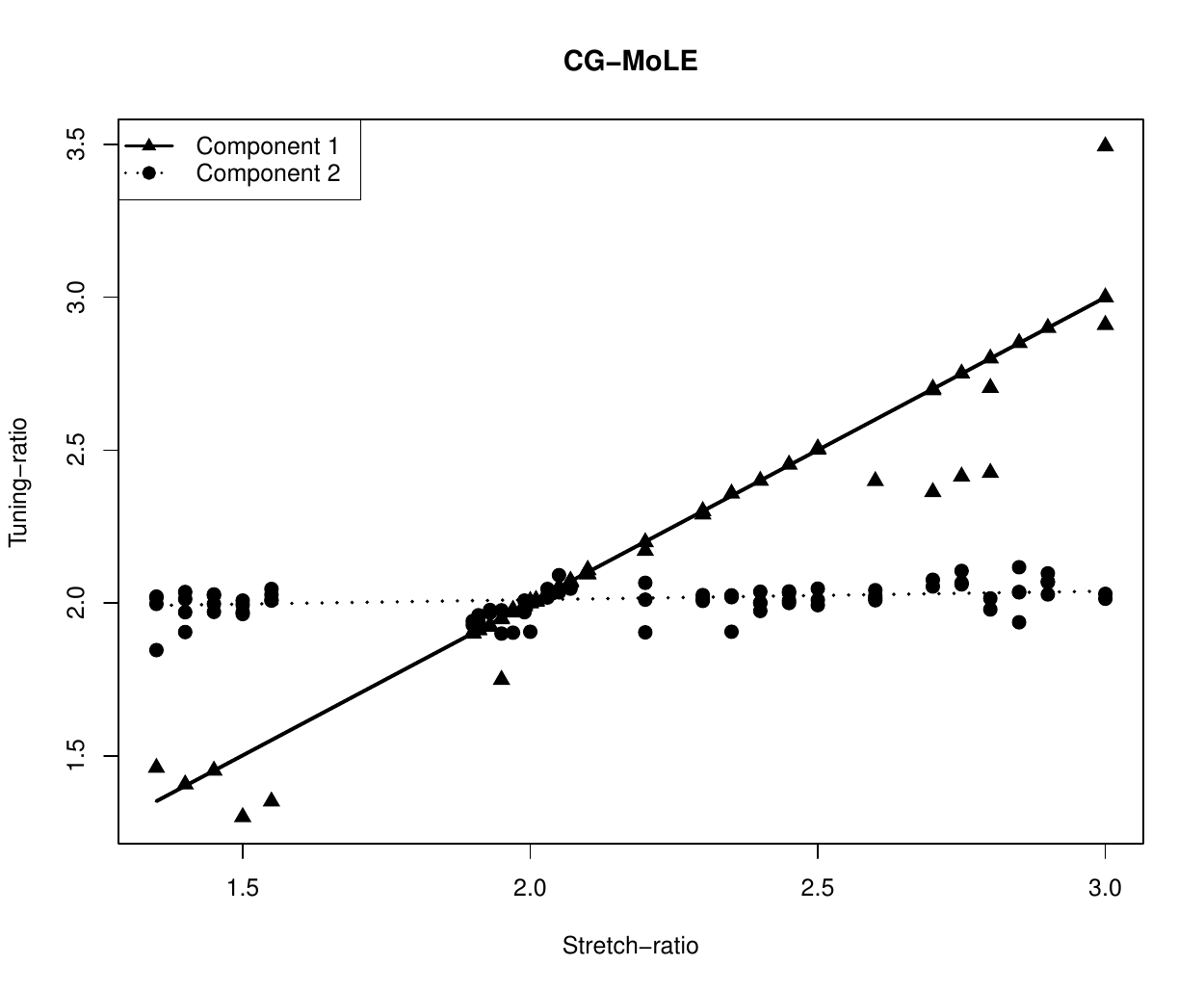}
		\caption{}
		\label{fig:second}
	\end{subfigure}
	\\
	\begin{subfigure}{0.47\textwidth}
		\centering
		\includegraphics[width=\textwidth]{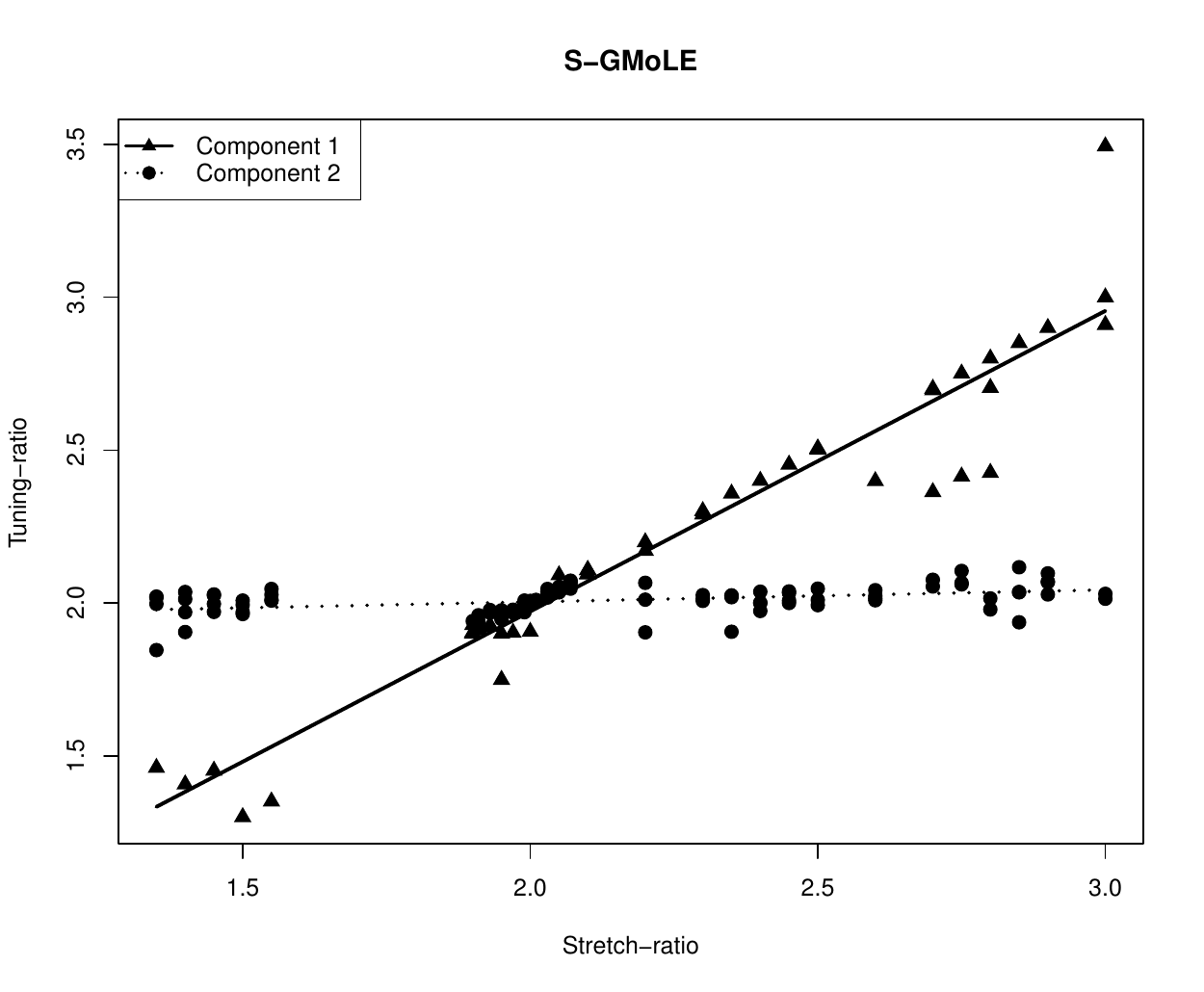}
		\caption{}
		\label{fig:first}
	\end{subfigure}
	\hfill
	\begin{subfigure}{0.47\textwidth}
		\centering
		\includegraphics[width=\textwidth]{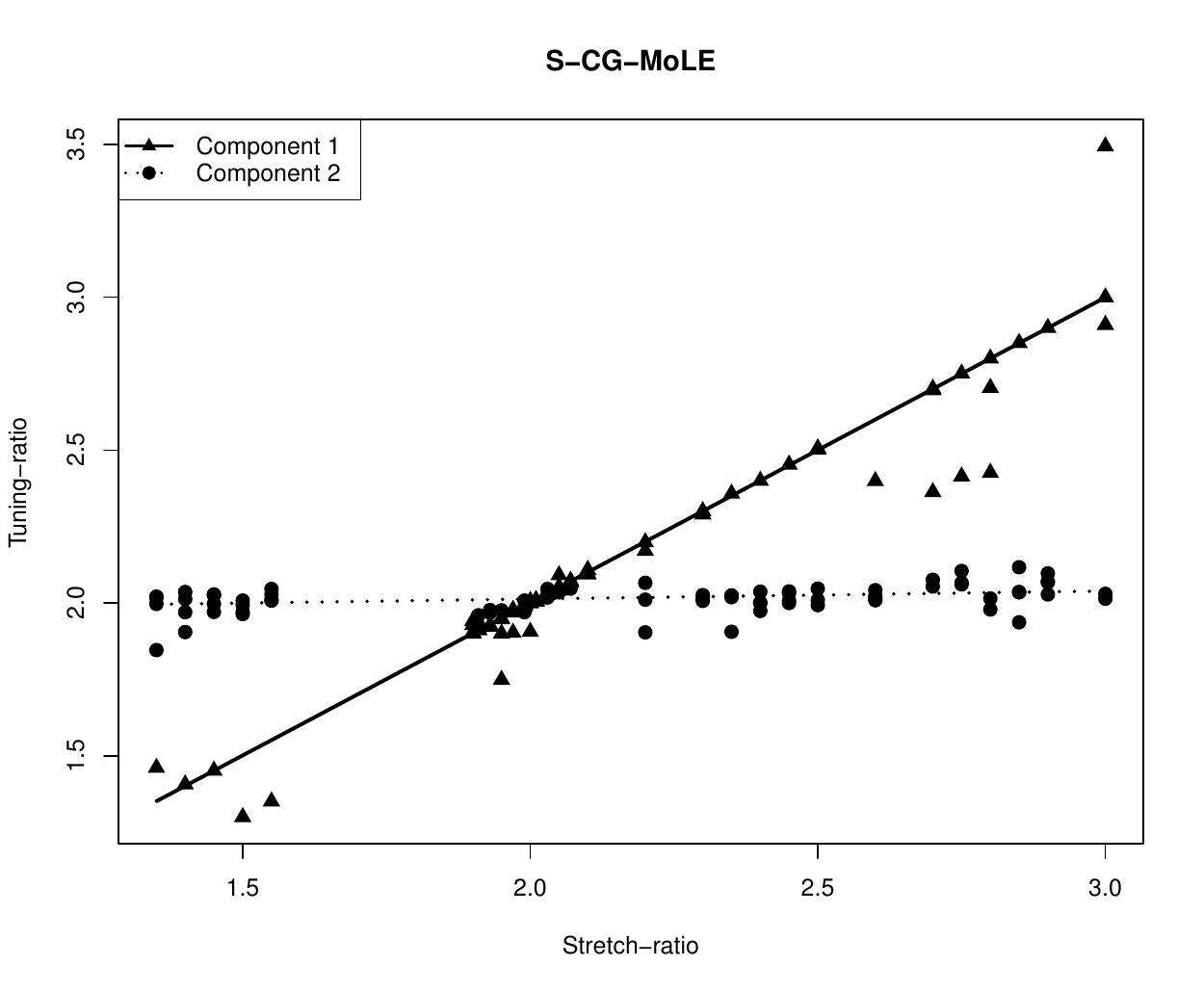}
		\caption{}
		\label{fig:second}
	\end{subfigure}
	\caption{Estimated models on the tone perception data: (a) GMLRs (b) CGMLRs (c) GMoE (d) CG-MoE (e) S-G-MoE (f) S-CG-MoE}
	\label{fitted_models1}
\end{figure}
\begin{figure}[H]
	\centering
	\includegraphics[width=0.75\textwidth]{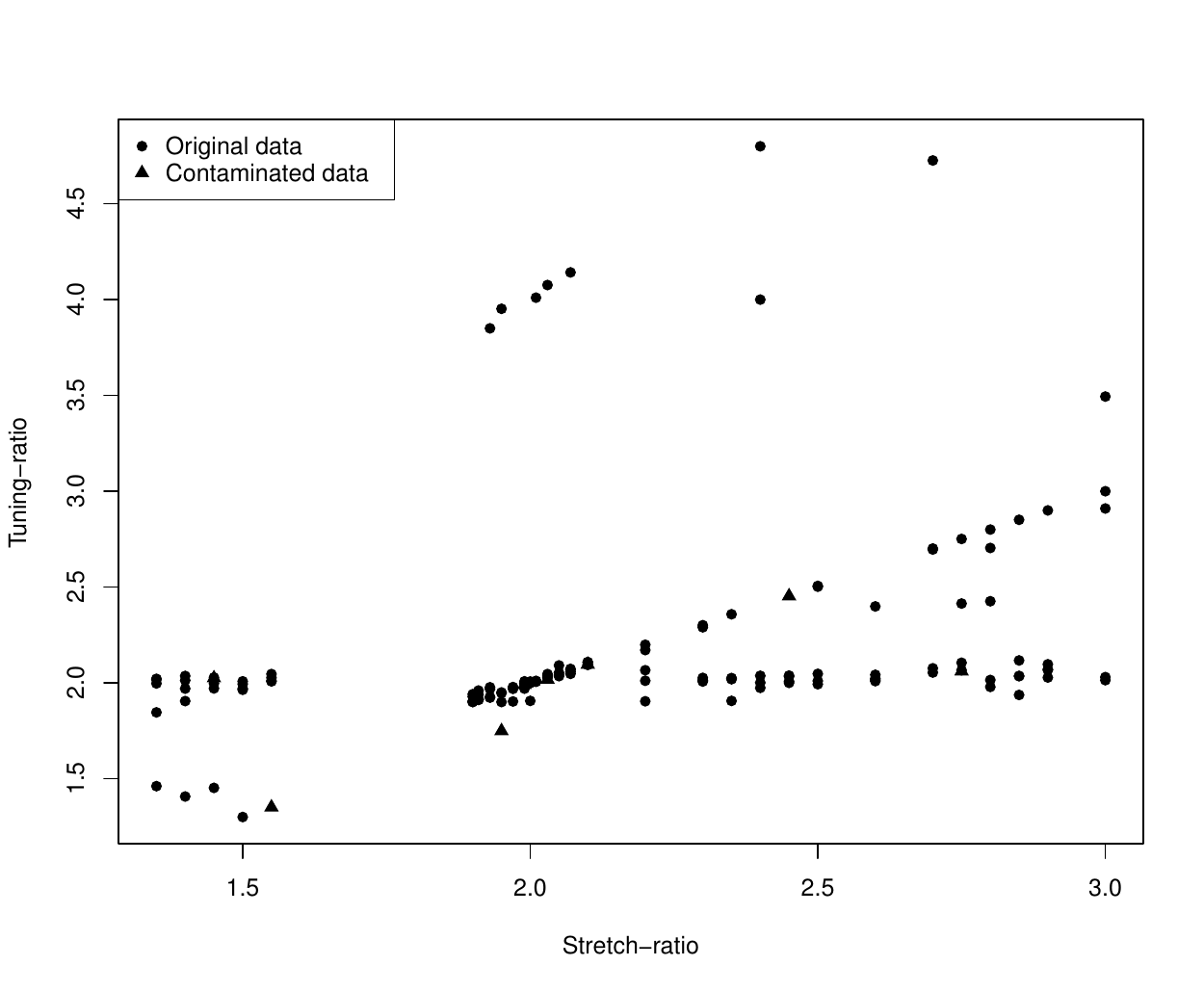}
	\caption{Scatter plot of the contaminated tone perception data}
	\label{tonedata2}
\end{figure}

Figure \ref{fitted_models1} plots the fitted regression lines obtained from all six fitted models. The component regression lines look virtually the same. In Figure \ref{fitted_mix1}, we plot the mixing proportions obtained from all the fitted models. The results are different across all the fitted models. The constant As expected, the parametric mixtures of experts (GMoLE and CGMoLE) give monotonically increasing functions for the mixing proportions. Notably, the semi-parametric mixtures of regressions (S-GMoLE and S-CG-MoLE) show opposing patterns at different values of the stretch-ratio. For the S-GMoLE model, when the stretch-ratio is between 1.5 and 2, the probability of belonging to component 1 decreases and then increases when the stretch-ratio is between 2 and 2.5. The opposite relationship is true for the S-CG-MoLE model. Overall, the contaminated mixture of regression models gives the best fit based on the BIC.\\
To evaluate the performance of the proposed methods in the presence of outliers, we add artificial noise to the tone perception data. More specifically, the data are contaminated by multiplying the $y-$values of $5\%$ of a random selection of data points by $2.5$. The resulting contaminated data is plotted in Figure \ref{tonedata2}.

\begin{figure}[H]
	\centering
	\includegraphics[width=0.75\linewidth]{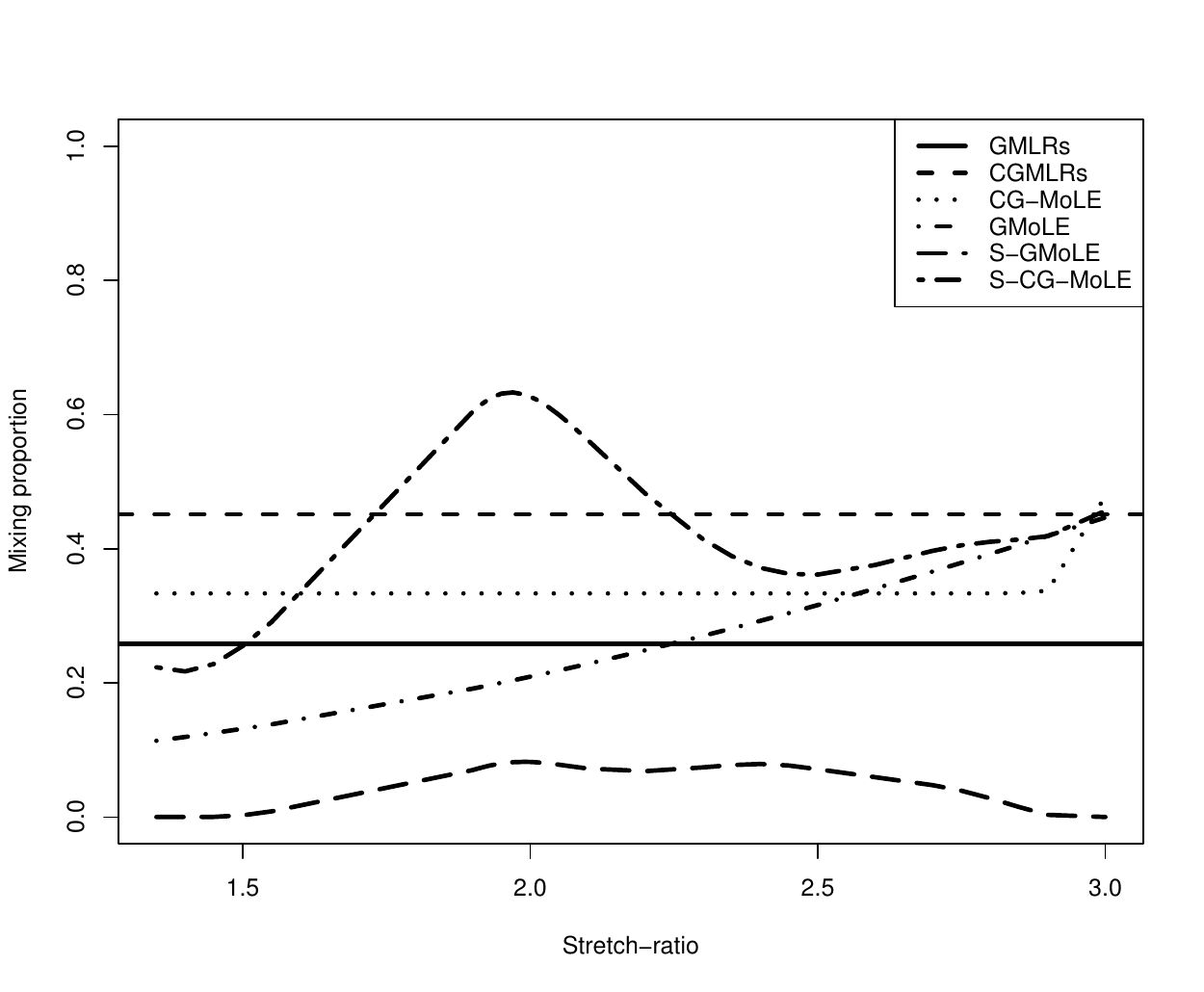}
	\caption{Estimated mixing proportions for the first component of all the six fitted models on the contaminated tone perception data}
	\label{fitted_mix2}
\end{figure}

\begin{figure}[!ht]
	\centering
	\begin{subfigure}{0.47\textwidth}
		\centering
		\includegraphics[width=\textwidth]{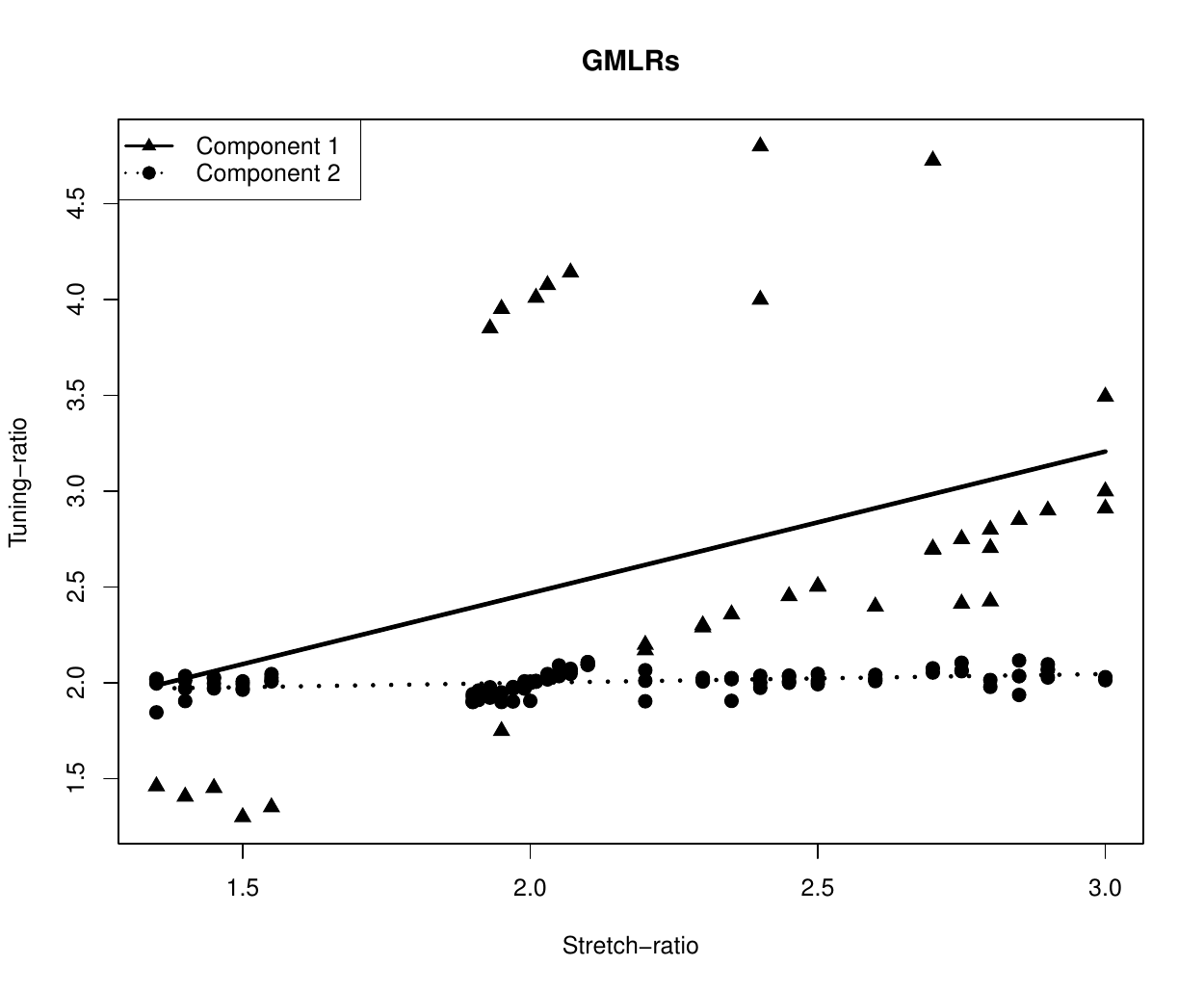}
		\caption{}
		\label{fig:first}
	\end{subfigure}
	\hfill
	\begin{subfigure}{0.47\textwidth}
		\centering
		\includegraphics[width=\textwidth]{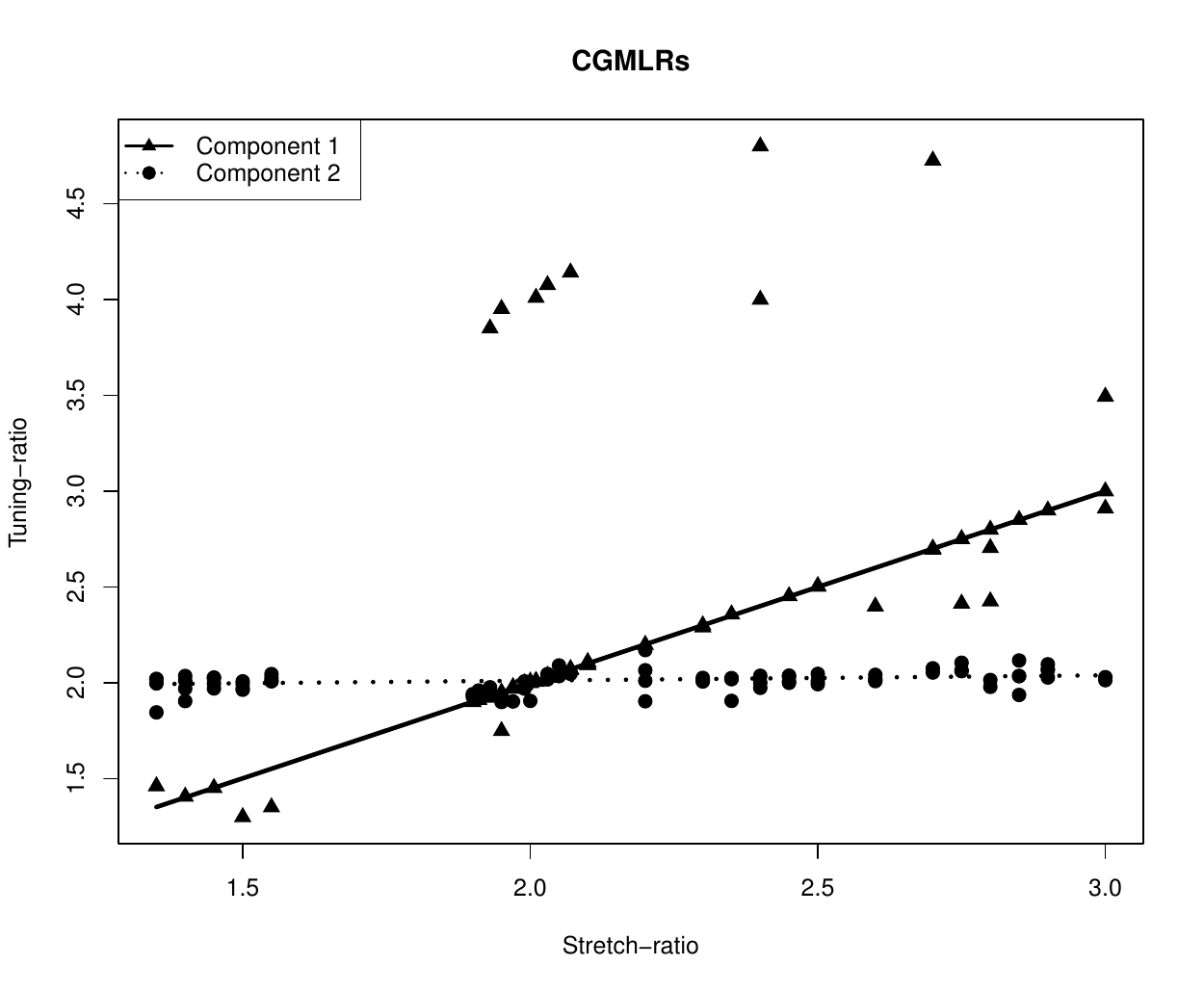}
		\caption{}
		\label{fig:second}
	\end{subfigure}
	\\
	\begin{subfigure}{0.47\textwidth}
		\centering
		\includegraphics[width=\textwidth]{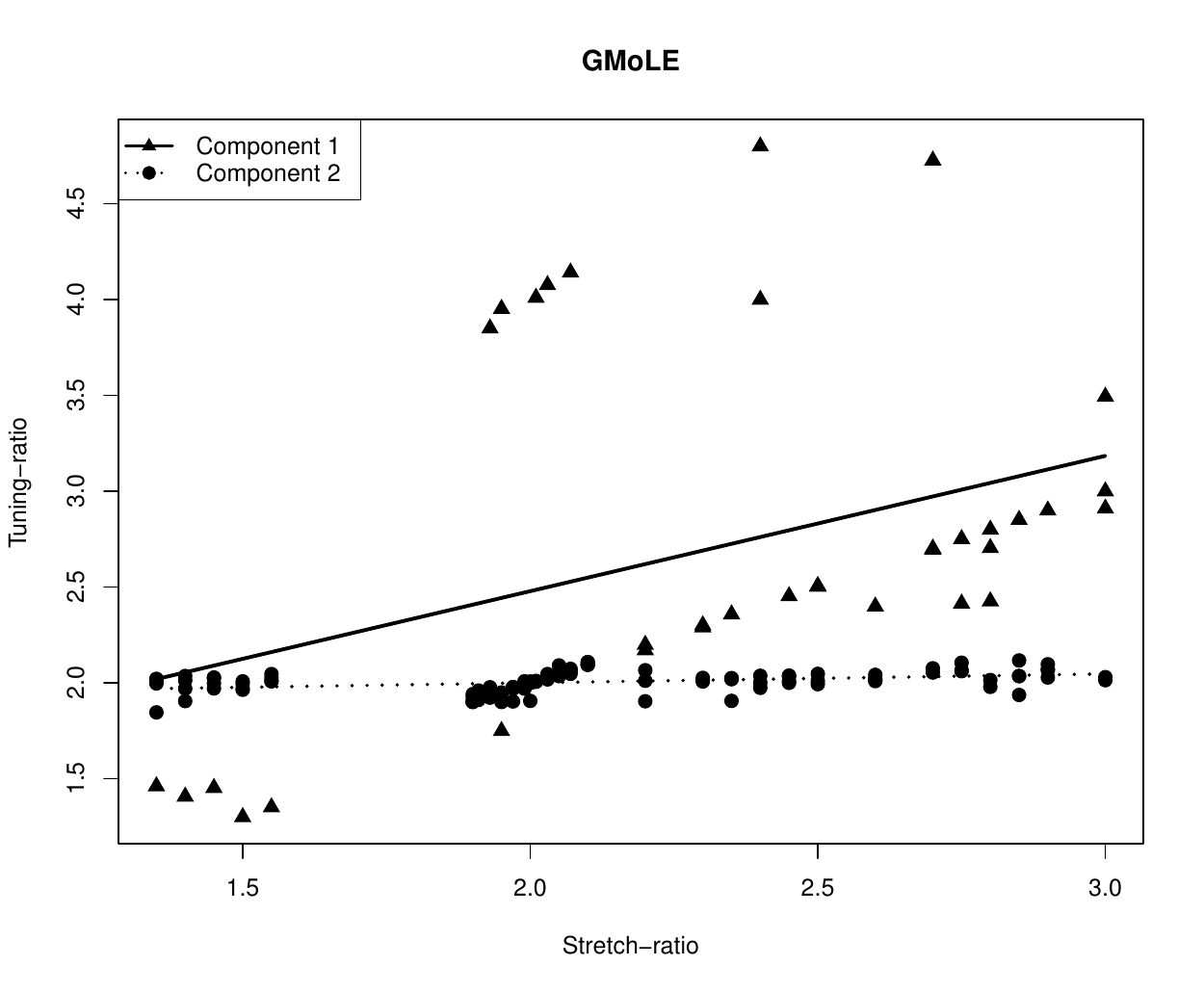}
		\caption{}
		\label{fig:first}
	\end{subfigure}
	\hfill
	\begin{subfigure}{0.47\textwidth}
		\centering
		\includegraphics[width=\textwidth]{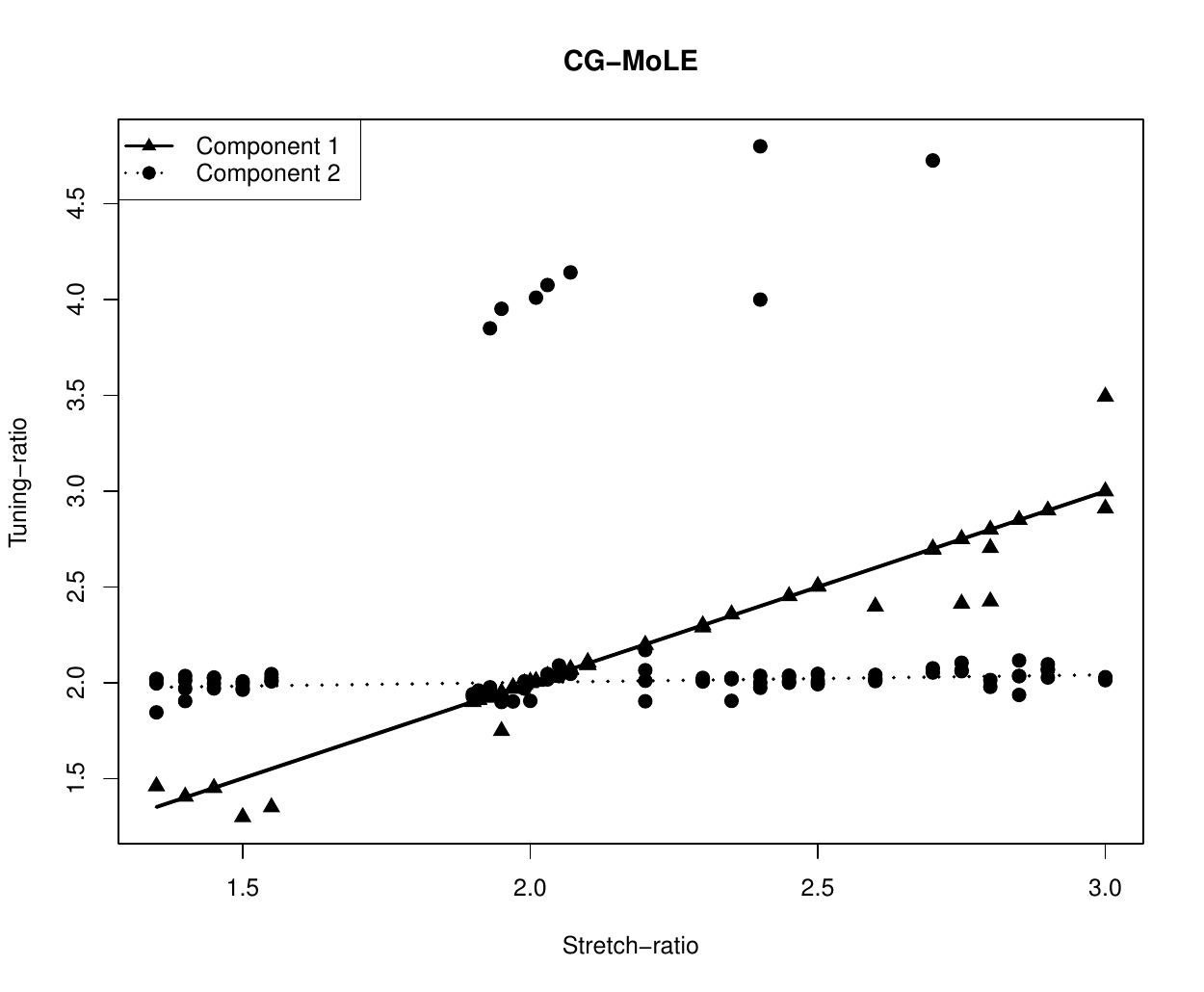}
		\caption{}
		\label{fig:second}
	\end{subfigure}
	\\
	\begin{subfigure}{0.47\textwidth}
		\centering
		\includegraphics[width=\textwidth]{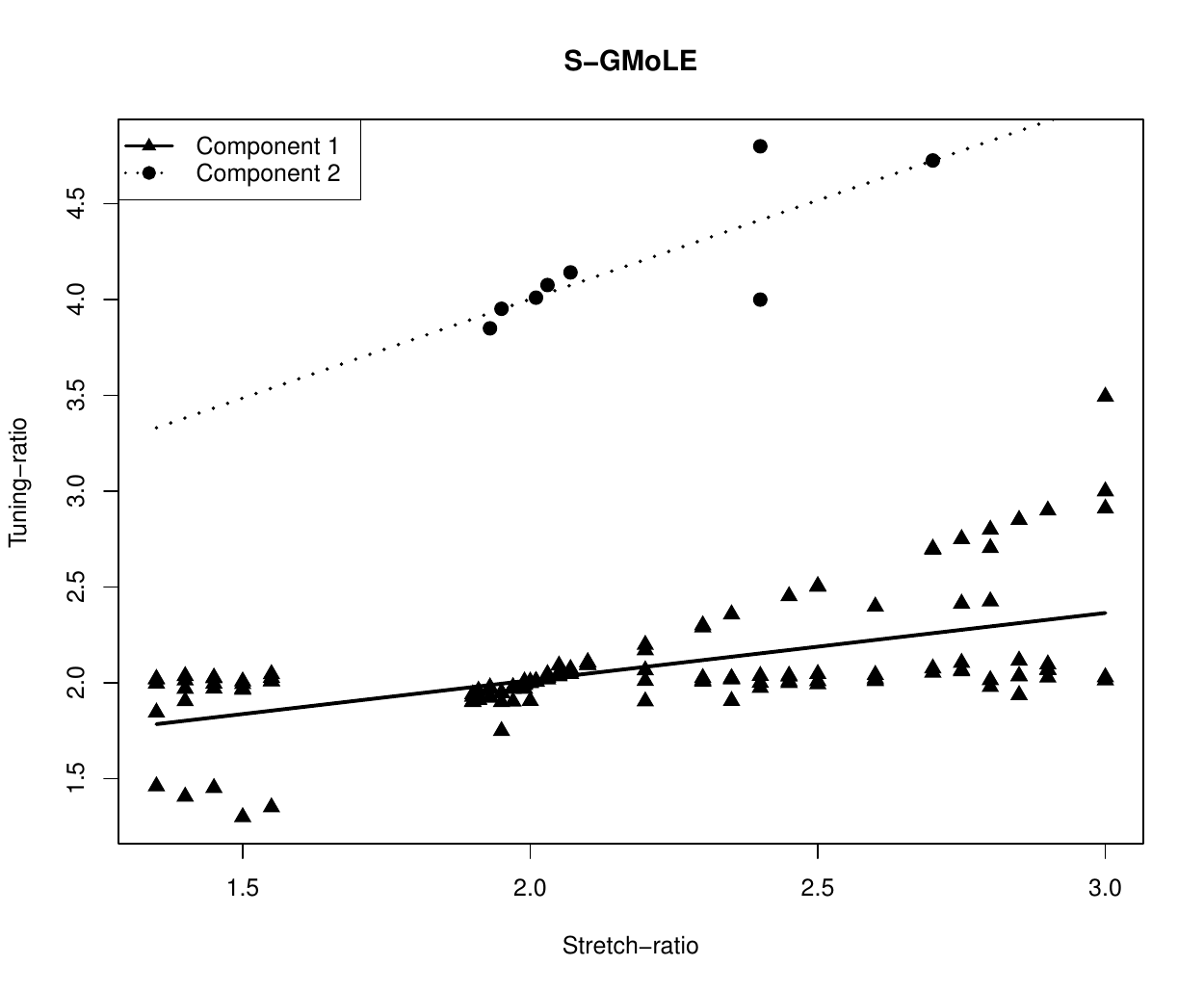}
		\caption{}
		\label{fig:first}
	\end{subfigure}
	\hfill
	\begin{subfigure}{0.47\textwidth}
		\centering
		\includegraphics[width=\textwidth]{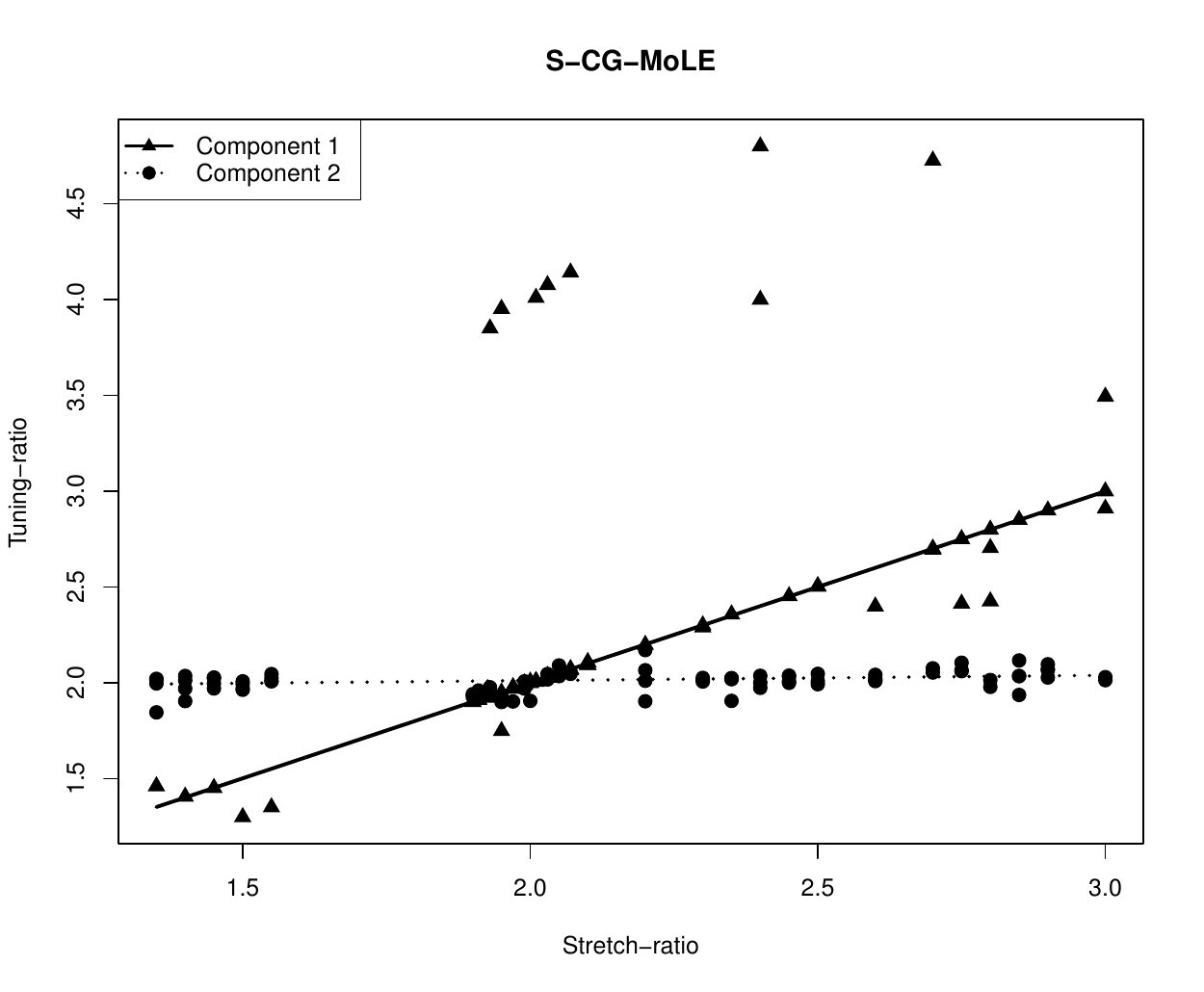}
		\caption{}
		\label{fig:second}
	\end{subfigure}
	\caption{Estimated models on the contaminated tone perception data: (a) GMLRs (b) CGMLRs (c) GMoE (d) CG-MoE (e) S-G-MoE (f) S-CG-MoE}
	\label{fitted_models2}
\end{figure}

We fit the six models to the contaminated tone perception data. The results are given in Table \ref{tab:fitted_models2}. The contaminated mixture of regression models (CGMLRs, CG-MoLE, and S-CG-MoLE) overwhelmingly outperform the Gaussian mixture of regression models (GMLRs, GMoLE, and S-GMoLE). This can also be seen in Figure \ref{fitted_models2}, which plots the component regression lines of all the fitted models. The left panel plots the models GMLRs, GMoLE, and S-GMoLE, and the right panel plots the models CGMLRs, CG-MoLE, and S-CG-MoLE. As can be seen from the figure, the former models are influenced by outliers, whereas the outliers seem to have no influence on the latter models. A comparison of Figures \ref{fitted_models1} and \ref{fitted_models2} shows that there is no difference in the fitted contaminated mixtures of regressions. This highlights the robustness of the contaminated models in the presence of outliers. Finally, in Figure \ref{fitted_mix2}, we plot all the fitted mixing proportions. Notably, the mixing proportions for all the contaminated models exhibit the same relationship as the ones obtained when the data has no outliers. 

\begin{table}[!ht]
\footnotesize
	\caption{Results of the models fitted on the contaminated tone perception data}
	\centering
	\begin{tabular}{|c|c|c|c|c|c|c|}
		\hline
		&\text{GMLRs}&\text{CGMLRs}&\text{GMoE}&\text{CG-MoE}&\text{S-G-MoE}&\text{S-CG-MoE}\\
		\hline
		$\beta_{10}$&0.9885&0.0040&1.0664&0.0041&1.3100&0.0040\\ 
		$\beta_{20}$&1.9090&1.9565&1.9071&1.9242&1.9368&1.9568\\
		$\beta_{11}$&0.7396&0.9986&0.7061&0.9985&0.3515&0.9985\\ 
		$\beta_{21}$&0.0457&0.0276&0.0465&0.0392&1.0327&0.0275\\
		$\pi_1$&0.2579&0.4513&-&-&-&-\\ 
		$\sigma_1$&0.8414&0.0042&0.8528&0.0042&0.0539&0.0042\\
		$\sigma_2$&0.0506&0.0284&0.0510&0.0463&0.0417&0.0238\\ 
		$\gamma_{10}$&-&-&-3.5595&114.3463&-&-\\ 
		$\gamma_{11}$&-&-&1.1150&-38.0553&-&-\\
		$\eta_1$&-&94535.212&-&69413.645&-&91350.613\\
		$\eta_2$&-&7.7132&-&1162.205&-&7.5836\\
		$\alpha_1$&-&0.7020&-&0.8311&-&0.7008\\
		$\alpha_2$&-&0.5301&-&0.8825&-&0.5330\\
		\hline
		BIC&-78&$\mathbf{-292.9539}$&-79.9801&-278.5077&355.4877&-203.4893\\
		\hline
	\end{tabular}
	\label{tab:fitted_models2}
\end{table}

\newpage
\section{Conclusion}\label{sec6}
In this paper, we introduce a robust mixture-of-experts model based on contaminated Gaussian experts (or components) and flexible gating functions (or mixing proportions). More specifically, the mixing proportion functions are assumed to be smooth, unknown, hence non-parametric, functions of the covariates, and each component is assumed to have a contaminated Gaussian distribution. Maximum likelihood estimation through the expectation conditional maximization (ECM) algorithm was proposed for estimating the model. The non-parametric mixing proportion functions are estimated using local-linear kernel estimators. The practical usefulness and robustness of the proposed methods in the presence of mild outliers were demonstrated using an extensive simulation study and an application on a real dataset.\\
In order to avoid the practical consequences of the curse-of-dimensionality in local-kernel estimation (that is, data sparsity in local regions as we increase the dimensionality), we assumed that the mixing proportion functions are functions of a one-dimensional covariate. However, the model can be easily extended to accommodate multi-dimensional covariates by making use of additive models or machine learning approaches such as a neural network \cite{xue2022} or a random forest. 


\section*{Declarations}

\begin{itemize}
	\item Conflict of interest/Competing interests: The authors declare no conflict of interest.
	\item Data availability: The data that support the findings of this study are publicly available.
\end{itemize}

\newpage
\bibliography{sn-bibliography}
\bibliographystyle{plainnat}

\end{document}